\documentclass{aa}

\usepackage{longtable,lscape}
\usepackage{graphicx}
\usepackage{txfonts}
\begin{document}

   \title{The Kormendy relation for early-type galaxies. Dependence on the magnitude range.}

   \titlerunning{The Kormendy relation. Dependence on the magnitude range.}

   \authorrunning{Nigoche-Netro {\it et al.}}


   \author{A. Nigoche-Netro\inst{1}, A. Ruelas-Mayorga\inst{2} \and A. Franco-Balderas \inst{2} 
          }


   \offprints{A. Nigoche-Netro}


   \institute{Instituto de Astrof\'isica de Canarias (IAC),
              V\'ia L\'actea s/n, 38200 La Laguna, Spain\\
              \email{anigoche@iac.es}
         \and
            Instituto de Astronom\'ia, Universidad Nacional Aut\'onoma de M\'exico. Apartado Postal 70-264, M\'exico D.F., C.P. 04510\\
            \email{rarm@astroscu.unam.mx; alfred@astroscu.unam.mx}
             }



  \abstract
   {}
   {Previous studies indicate that faint and bright early-type galaxies (ETGs) present different coefficients and dispersion for their Kormendy relation (KR). A recently published paper states that the intrinsic dispersion of the KR depends on the magnitude range within which the galaxies are contained, therefore, we investigate here whether the magnitude range has also an influence over the values of the coefficients of the KR; $\alpha$ (zero point) and $\beta$ (slope). If the values of the KR coefficients depend on the magnitude range, and this fact is not considered when performing comparisons of different galaxy samples, the differences which might be found may be misinterpreted.}
   {We perform an analysis of the KR coefficients for 4 samples of galaxies, which contain an approximate total of 9400 ETGs in a relatively ample magnitude range ($<\Delta M>$ $\sim 6$ $mag$). 
We calculate the values of the KR coefficients in two ways: i) We
consider the faintest galaxies in each sample and we progressively increase the width of the magnitude interval by inclusion of the brighter galaxies (increasing magnitude intervals), and ii) we consider narrow magnitude intervals of the same width ($\Delta M = 1.0$ $mag$) over the whole magnitude spectrum available (narrow magnitude intervals). We also perform simulations of the distribution of galaxies in the log($r_{e}$) - $<\mu>_{e}$ plane and compare the KR defined by the simulations with that obtained from the real galaxy samples.}
   {The main results we find are as follows: i) In both increasing and narrow magnitude intervals the KR coefficients change systematically as we consider brighter galaxies, ii) non-parametric tests show that the fluctuations in the values of slope of the KR are not products of chance variations and that there is evidence of an underlying trend, and iii) this trend suggest a maximum of the slope around absolute magnitude $M_{B}\sim -18\pm1$.}
   {The analysis of the results makes us conclude that the values of the KR coefficients depend on the width of the magnitude range and the brightness of galaxies within the magnitude range. This dependence is due to the fact that the distribution of galaxies in the $\log (r_{e}) - <\mu>_{e}$ plane depends on luminosity and that this distribution is not symmetrical, that is, the geometric shape of the distribution of galaxies in the $\log (r_{e}) - <\mu>_{e}$ plane plays an important role in the determination of the values of the coefficients of the
KR.}


   \maketitle

%

\section{Introduction}

\begin{table*}



 \centering
 \begin{minipage}{170mm}
\caption{Name, Number of galaxies, Magnitude range (according to the papers where the samples were taken from), Approximate magnitude range in the B-Filter (calculated by us), redshift and type of photometric profile (TP) (S = S\'ersic, dV = de Vaucouleurs) of the different galaxy samples compiled in this work. }

  \begin{tabular}{@{}lccccrrr@{}}
\hline

Sample   & N & Magnitude range & Approximate magnitude range in the B-filter & $z$ &TP \\

 \hline
 
 7 Abell clusters (V filter) & 626 & $-16.0 \ge M_{V} > -22.0$ & $-15.1 \ge M_{B}> -21.1$ & $0.048$ & S \\
 Coma cluster (Gunn r filter)& 196 & $-17.0 \ge M_{Gr}> -24.0$  & $-15.9 \ge M_{B}> -22.9$ & $0.024$ & S, dV\\
 Hydra cluster (Gunn r filter)& 54 & $-18.0 \ge M_{Gr}> -22.0$  & $-16.9 \ge M_{B}> -20.9$  & $0.014$ & dV\\
 SDSS (g* filter) & 8666  & $-18.0 \ge M_{g*}> -24.1$  & $-17.5 \ge M_{B}> -23.6$ & $\le 0.3$ & dV\\
 SDSS (r* filter) & 8666  &  $-18.6 \ge M_{r*}> -24.7$  & $-17.5 \ge M_{B}> -23.6$ & $\le 0.3$ &dV\\
 SDSS (i* filter) & 8666  &  $-19.0 \ge M_{i*}> -25.1$  & $-17.5 \ge M_{B}> -23.6$  & $\le 0.3$ &dV\\
 SDSS (z* filter) & 8666  & $-19.3 \ge M_{z*}> -25.3$  & $-17.5 \ge M_{B}> -23.6$  & $\le 0.3$ &dV\\

\hline
\end{tabular}
\end{minipage}



\end{table*}

It is a well known fact that the structural parameters of ordinary ETGs follow the Fundamental Plane (FP) relation (\cite{djo87,dre87}). The FP relation is usually expressed as a correlation among the {\it logarithm of the effective radius} ($\log{\kern 1pt} (r_{e} )$), {\it effective mean} {\it surface brightness} ($<\mu>_{e}$) and {\it logarithm of the central velocity dispersion} (log ($\sigma_{0}$)), and is expressed mathematically with the following equation:

\begin{equation}
\log\,(r_{e})\, \; =\, \; a\,\log\,(\sigma_{0})\, \; +\; b\, \, <\mu>_{e} \;
+\; c
\end{equation}

The FP relation is a direct consequence of the dynamical equilibrium
condition (virial theorem) and of the regular behaviour of both the
mass-luminosity ratio and of the galactic structure of the ETGs along all the luminosity range. Due to its small intrinsic dispersion ($\sim0.1$ $dex$ in $r_{e}$ and $\sigma _{0}$ and $\sim0.1$ $mag$ in $<\mu>_{e}$) the FP is considered a powerful tool in measuring galactic distances and also in studies of galactic formation and evolution (\cite{kja93,jor96,jor99,kel97}).

A physically significant projection of the FP is the correlation between $\log{\kern 1pt} (r_{e})$ and $<\mu>_{e}$, known as the Kormendy relation (KR):

\begin{equation}
<\mu>_{e}\, \; =\, \; \alpha\; +\; \beta\, \log(r_{e})
\end{equation}

Several studies demonstrate that the ETGs in clusters define the KR with an intrinsic dispersion of approximately 0.4 $mag$ in $<\mu>_{e}$ (\cite{ham87,hoe87,san91,san01,lab03}). Some authors contend that the high intrinsic dispersion reported is due mainly to the fact that the KR does not consider the third parameter (the velocity dispersion of the FP)  (Ziegler {\it et al.} 1999), besides the value of this dispersion must be slightly increased by the measurement errors as well as the systematic errors due to the photometry calibration and also due to the corrections introduced for different biases (zero point and color transformation, K-correction and reddening).

Recent studies show that low-density-environment ETGs and isolated ETGs also follow the KR with the same coefficients and intrinsic dispersion as do ETGs in clusters (\cite{red04,nig07}). On the other hand, previous studies demonstrated that ETGs (plus bulges of spiral galaxies) formed two distinct families on the structural parameters plane; i.e. the family of bright ETGs ($M_{B}\leq-18$) that follows the KR and the family of dwarf ETGs ($M_{B}>-18$) with more disperse and heterogeneous properties (their effective parameters range over a wide interval for the same total luminosity) (\cite{kor85,cap92}). Subsequently Graham \& Guzm\'{a}n (2003) showed that the difference between the bright elliptical (bright E) and the dwarf elliptical (dE) galaxies is only apparent and that there is a continuous structural relation between both classes. They say that the different behaviour presented by the bright E and the dE galaxies in the log($r_{e}$) - $<\mu>_{e}$ plane and the variations in the value of the slope of the relation do not imply a different formation mechanism, rather it may be interpreted as a systematic change in the shape of the light-profile with galactic magnitude ($M$). This result, together with the fact that the intrinsic dispersion of the KR depends on the magnitude range (\cite{nig07,nig07b}) prompted us to study the behaviour of the coefficients of the KR as functions of the magnitude range. If indeed, the values of the coefficients of the KR depend on the magnitude range within which the galaxies are contained, and this fact is not considered when performing comparisons of galaxy samples such as the dependence of the KR on the environment, on redshift or on wavelength, the differences which might be found may be misinterpreted.

In this paper, we present a compilation of 4 samples of ETGs which in total contain approximately 9400 galaxies and cover a relatively ample magnitude range ($<\Delta M>$ $\sim 6$ $mag$). Using these data we analyze the behaviour of the coefficients of the KR with respect to several characteristics of the magnitude range. We also present simulations of the distribution of the galaxies on the log($r_{e}$) - $<\mu>_{e}$ plane. These simulations allow us to reproduce in a reasonable manner the results we obtain from the real samples of galaxies.

This paper is organized as follows. In section 2, we present the different samples used in the analysis of the KR. Section 3 describes the fitting method used in calculating the KR coefficients, as well as the behaviour of the $\beta$ coefficient with respect to the absolute magnitude range. Section 4 presents simulations of the distribution of the galaxies on the log($r_{e}$) - $<\mu>_{e}$ plane and finally in section 5 the conclusions are presented.

\section{The samples}


\begin{figure*}

\centering

\includegraphics[bb= 20 20 750 750,angle=-90,width=12cm,clip]
{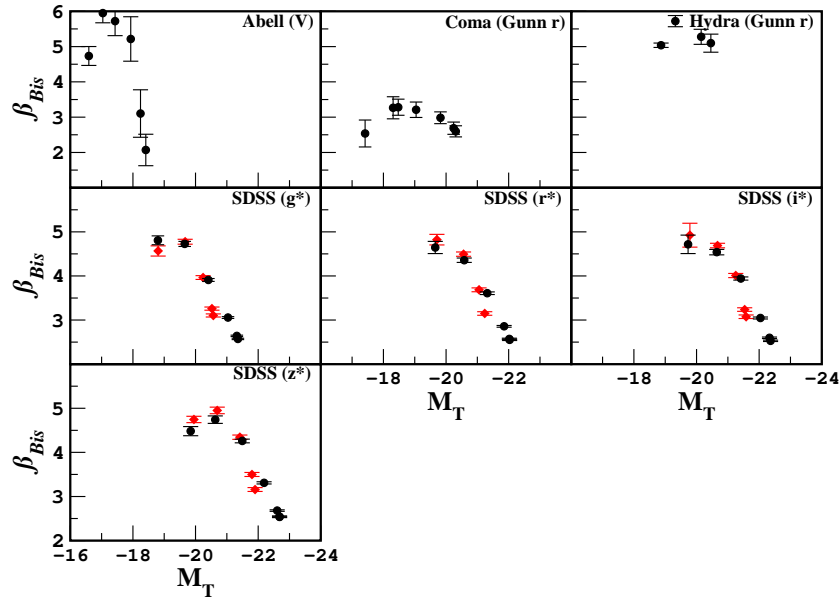}

 \vspace*{-90pt}

  \caption{Variation of the KR slope ($\protect\beta$) for the different samples of galaxies in increasing magnitude intervals (upper magnitude cut-off). This graphs
shows the $\protect\beta$ coefficient obtained using the $BCES_{Bis}$
method. Each point corresponds to the mean value of the total absolute magnitude of the galaxies contained in each magnitude interval (see table A.1). Diamonds represent the SDSS homogeneous sample.}

\end{figure*}


We use a Sloan Digital Sky Survey (SDSS) sample of 8666 ETGs (\cite{ber03}) in filters g*, r*, i* and z* (total absolute magnitude range $-18 \ge M_{g^{*}} > -24.1$ and its equivalent in other filters) as well as a sample of 626 ETGs in the Johnson V filter ($-16 \ge M_{V}> -22$) from 7 Abell clusters (WINGS project, \cite{fas02,var04}), a sample in filter Gunn r ($-17 \ge M_{Gr} > -24$) with 196 ETGs from the Coma cluster (\cite{jor95,mil97,jor99a,agr05}) and a 54 ETGs sample in the filter Gunn r ($-18 \ge M_{Gr}> -22$) from the Hydra cluster (\cite{mil97}). In the 7 Abell clusters sample we include the following clusters: A147, A168, A193, A2457, A2589, A2593, and A2626. All the samples are redshift-homogeneous (the galaxies are contained within a narrow redshift interval), except for the SDSS sample which cover a relatively ample redshift interval (0.01 $\leq\;z\;\leq$ 0.3). This sample is magnitude-limited (\cite{ber03b}). Besides, within large volumes there could be evolution effects of the parameters of the galaxies. So, in order to have a representative sample of the universe in a given volume without any evolution effects it is important to consider narrow redshift intervals. Bernardi {\it et al.} 2003b recommend $\Delta$$z$ = 0.04. This value comes from the sizes of the largest structures in the universe seen in numerical simulations of the cold dark matter family of models (\cite{col00}). On the other hand, it is also well known that the SDSS photometry underestimates the luminosity of the brightest objects in crowded fields (\cite{ber07,ber07b}). To probe the possible evolution effects and the photometric bias of the brightest galaxies, we have built a subsample from the SDSS in the redshift interval 0.04 $\leq\;z\;\leq$ 0.08. This subsample has 1670 galaxies in each filter and cover a magnitude range $<\Delta M>$ $\sim 4$ $mag$ ($-18.5 \ge M_{g^{*}} > -22.9$ and its equivalent in other filters). This subsample will be referred to, from now on, as homogeneous sample from the SDSS.

In Table 1 we present relevant information (number of galaxies, magnitude range, redshift and type of photometric profile)
for the samples of galaxies we use in this paper. The magnitude range information is given in relation to the different filters used (from the literature) and also the approximate range for the B-magnitude (calculated by us). The transformation to the B filter was accomplished by use of the following equations: B - Gunn r = 1.15 (\cite{mil97}), B - V = 0.92 (\cite{mic00}) and B - g* = 0.5 (\cite{fuk96}).

All the samples consider the photometric parameters log($r_{e}$) and 
$<\mu>_{e}$ corrected for different biases, such as: seeing (Saglia {\it et al.} 1993), galactic extinction (Schlegel {\it et al.} 1998), K correction (Jorgensen {\it et al.} 1992; Bruzual \& Charlot 2003) and cosmological dimming (Jorgensen {\it et al.} 1995). These parameters as well as their uncertainties were taken directly from the different papers cited above, except for the parameters for the Abell samples which constitute a private communication from the WINGS project team (\cite{fas02,var04}), and those for Coma (data from \cite{agr05}). In this last case, Aguerri {\it et al.} (2005) give information for effective surface brightness ($\mu_{e}$) instead of effective mean surface brightness ($<\mu>_{e})$, therefore, we transform these data following the procedure described by Graham \& Driver (2005). Additionally, to be consistent, we review carefully that the photometric parameters and their errors were obtained by the same methods for all samples.

It is important to note that, the photometric parameters of the faint and bright ETGs in the Abell sample as well as the faint ETGs in the Coma sample were obtained using S\'ersic profile-fits, whereas for the bright ETGs from the other samples (which for the SDSS sample consists of the entire sample), these parameters were obtained using  
de Vaucouleurs $r^{1/4}$ profile-fits (see Table 1). In the literature we find that the use of one profile or the other may affect significantly the estimations of the photometric parameters (Caon {\it et al.} 1993; Fritz {\it et al.} 2005), this in turn will also affect the estimations of the KR parameters (Ziegler {\it et al.} 1999; Kelson {\it et al.} 2000), however, we also find that the de Vaucouleurs profile-fits represent a good approximation to the S\'ersic profile-fits for bright galaxies (Prugniel \& Siemen 1997). In section 3.3 it will be shown that the use of the de Vaucouleurs profiles as approximations to the S\'ersic profiles for bright ETGs does not produce important biases to the results obtained in this paper.

Finally, an important characteristic of S0 galaxies is that, in general, the structural properties of their bulges show approximately the same homogeneity as those of E galaxies. Due to this fact, and because KR uses effective parameters, we use the photometric parameters from the whole galaxy in the case of bright Es, and from the bulge in the case of bright S0s. For dwarf galaxies we follow 
the same procedure: information from the whole galaxy for the dEs and from the bulge for the dS0s.

\section{Kormendy relation}

\subsection{Calculation of the Kormendy relation}

The estimation of the KR coefficients may be severely affected by the
fitting method and by the choice of dependent variable. The biases may be larger if there are measurement errors in the variables, if these errors are correlated and/or if there is intrinsic dispersion. The {\it Bivariate Correlated Errors and Intrinsic Scatter bisector} ($BCES_{Bis}$) fit (\cite{iso90,akr96}) is a statistical model that takes into account the different sources of bias mentioned above, which are precisely those that affect our samples. In this work we use the $BCES_{Bis}$ method for the determination of the KR coefficients.

From the photometric parameters of the different galaxy samples, we
calculate the coefficients of the KR in different magnitude ranges, the uncertainties of the coefficients, the correlation coefficient (Pearson Statistics) and the intrinsic dispersion of the KR (subtracting in quadrature from the intrinsic dispersion in $<\mu>_{e}$ the residues dispersion due to the measurement errors of $<\mu>_{e}$ and log($r_{e}$)) (\cite{lab03}). According to La Barbera {\it et al.} (2003), for the calculation of the intrinsic dispersion it is necessary to have the measurement errors in $<\mu>_{e}$ and log($r_{e}$), these errors come directly from the papers from which we take the galaxy samples (see section 2), while the errors in the KR coefficients were calculated by us following Akritas \& Bershady (1996) in 1$\sigma$ intervals. 


\begin{figure*}

\centering

\includegraphics[bb= 20 20 750 750,angle=-90,width=12cm,clip]
{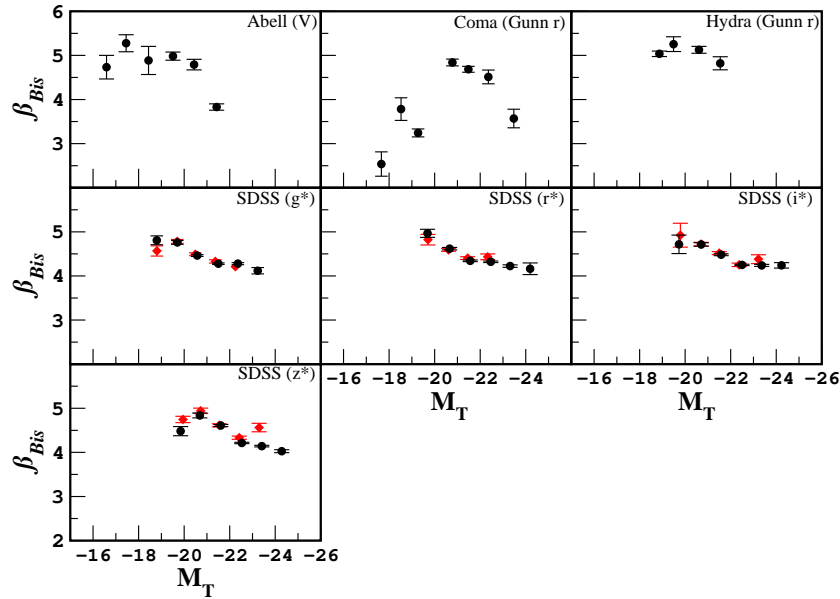}

 \vspace*{-90pt}

  \caption{Variation of the KR slope ($\protect\beta$) for the different samples of galaxies. The graphs show the $\protect\beta$ coefficient values obtained with the $BCES_{Bis}$ method. Each point represent a $1$ $mag$ interval (mean value of the total absolute magnitude of the galaxies contained in each magnitude interval analyzed, see table A.3). Diamonds represent the SDSS homogeneous sample.}

\end{figure*}


The KR coefficients were calculated both in increasing
magnitude intervals, as well as, in narrow magnitude intervals.
This allows us, among other things, to characterize the behavior of the KR coefficients with respect to the width of the magnitude range and the brightness of galaxies within the magnitude range, so the results obtained may be utilized as reference for other studies in which different magnitude ranges are used. In appendices A and B
we present the results for the values of the coefficients of the KR for the different samples of galaxies and for different magnitude intervals.

Here it is necessary to mention that both the figures in the main section of this paper as well as those tables in appendix A, contain the photometric information given in the filters in which the samples were originally observed, however for comparison purposes of the values of the KR coefficients for different samples, we present in appendix B the relevant figures (Figs. 1, 2 and 7) and the corresponding tables with the photometric information given in the B-filter reference frame.

\subsection{Behaviour of the Kormendy relation coefficients with respect to absolute magnitude range}

From the analysis of the data for our different samples we notice that the intrinsic dispersion of the KR ($\sigma_{KR}$), as clearly stated by Nigoche-Netro {\it et al.} (2007), changes appreciably each time we include brighter galaxies in the samples (upper magnitude cut-off), that is to say when we use increasing magnitude intervals (see Tables A.1 and B.1). We also notice that the correlation coefficient (R) for each fit diminishes considerably. Apart from these changes, we can also see changes in the coefficients of the KR and that these changes are larger than the associated errors for most of the cases 
(differences in the $\beta$ coefficient may be as large as 65\%).
The distribution of the $\beta$ coefficient may be seeing on Fig. 1 (see also Fig. B.1). On the other hand, if we perform the KR 
analysis considering first the brightest galaxies in each sample and we include progressively the fainter galaxies (lower magnitude cut-off) (see Tables A.2 and B.2), the behaviour of the KR parameters is similar to that described above: both the intrinsic dispersion and the KR coefficient-values change systematically as we increase the width of the magnitude interval. 

We also perform the analysis of the data using magnitude intervals that are narrow (see Tables A.3 and B.3), that is, considering galaxy samples in magnitude intervals of the same width and progressively brighter. For this case we can also see changes in the coefficients of the KR, however, the changes are less pronounced but still larger than the associated errors for most of the cases (differences in the $\beta$ coefficients may be as large as 48\%). We also find that the intrinsic dispersion of the KR is relatively low in all cases and that the correlation coefficient is, on the average, superior to
0.9. The variation of the $\beta$ coefficient when we consider narrow
magnitude intervals may be seen on Fig. 2 (see also Fig. B.2) where a dependence
between the $\beta$ coefficient and absolute magnitude appears to be hinted at (it is interesting to note that this distribution, in the case of the Coma and Abell clusters, seems to reach a maximum at $
M_{B}\sim -18\pm1$). However, it may be possible for the $\beta$ coefficient to be constant ($\beta$ = 5) and that the differences we find are the result of statistical fluctuations.

The question that $\beta$ be constant and equal to 5 comes from the
definition of absolute magnitude in terms of effective radius and effective mean surface brightness, that is:

\begin{equation}
M\; = \, <\mu>_{e}\; - \; 5\, \log\,(R_{e})\, - \; 2.5 \, \log\,(2\, \pi)\, \;
-\; 5\, \log\, (D)\, +\; 5
\end{equation}

where $R_{e}$ [arcsec] and $D$ [pc] represent the effective radius and the distance to the object in question respectively and $<\mu>_{e}$ $[mag/arcsec^{2}]$ is the effective mean surface brightness.

If we consider the effective radius in kiloparsecs ($r_{e}$) and a constant magnitude, we obtain the KR as follows:

\begin{equation}
<\mu>_{e}\, \; =\, \; \alpha\; +\; 5 \, \log(r_{e})
\end{equation}

Which implies that

\begin{equation}
\beta\; = \, 5
\end{equation}

However, the observational data for $\beta$ seem to move away from the expected value. In order to clear this point, it is necessary to apply non-parametric tests to make certain that the fluctuations in the values of the slope are not products of chance variations.

\subsection{Hypothesis Tests for the evaluation of $\beta$ data}

{\tiny
\begin{table*}
 \centering
 \begin{minipage}{100mm}

\caption{Hypothesis Test for the evaluation of the $\beta_{Bis}$ data from the different samples of galaxies. Tests 1, 2 and 3 correspond to the mean value, run and chi-square tests respectively. The percentages refer to the level of confidence with which we can reject the null hypothesis. }

  \begin{tabular}{@{}lccccr@{}}
  
\hline
&&&\\

 & Test 1 & Test 2 & Test 3 \\


&&&\\

Sample (increasing magnitude intervals)  &  & & \\

 \hline
&&&\\

 Abell  &94 \% &81 \% & ...  \\
 Coma  & 99 \% & 89 \% & ...  \\
 Abell+Coma+Hydra & 99 \% & ... & 95 \%  \\
 SDSS in each filter & 99 \% & 96 \% & ... \\
 SDSS sum of 4 filters & 99 \% & 96 \% & 90 \% \\
 Abell+Coma+Hydra+SDSS sum of 4 filters  & 99 \% & 96 \% & 95 \%  \\

&&&\\

 Sample (1 $mag$ interval)  &  & & \\

 \hline
&&&\\

 Abell  &84 \% &81 \% & ...  \\
 Coma  & 99 \% & 89 \% & ...  \\
 Abell+Coma+Hydra & 99 \% & ... & 95 \%  \\
 SDSS in each filter  & 99 \% & 96 \% & ... \\
 SDSS sum of 4 filters & 99 \% & 96 \% & 90 \% \\
 Abell+Coma+Hydra+SDSS sum of 4 filters & 99 \% & 96 \% & 95 \%  \\

\hline

\end{tabular}
\end{minipage}
\end{table*}

}


\begin{figure*}

\centering

\includegraphics[bb= 20 20 700 700,angle=-90,width=12cm,clip]{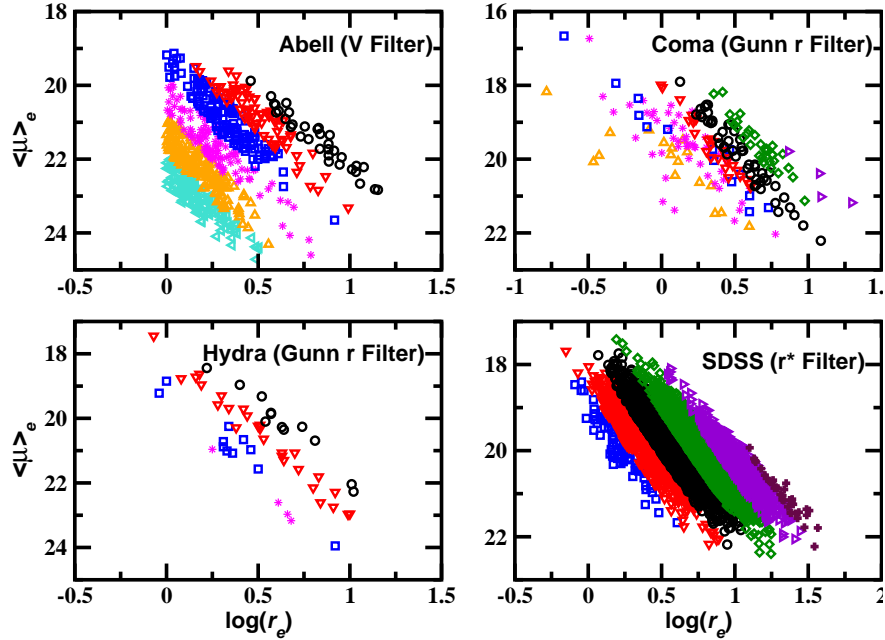}

 \vspace*{-70pt}

  \caption{Distribution of the Abell, Coma, Hydra and SDSS (r* filter) galaxies on the $log(r_{e}) - <\protect\mu>_{e}$ plane. Each symbol represents a $1$ $mag$ interval. The circles represent the (-21, -22] $mag$ interval.}

\end{figure*}



\begin{figure*}

\centering

\includegraphics[bb= 20 20 700 720,angle=-90,width=8cm,clip]{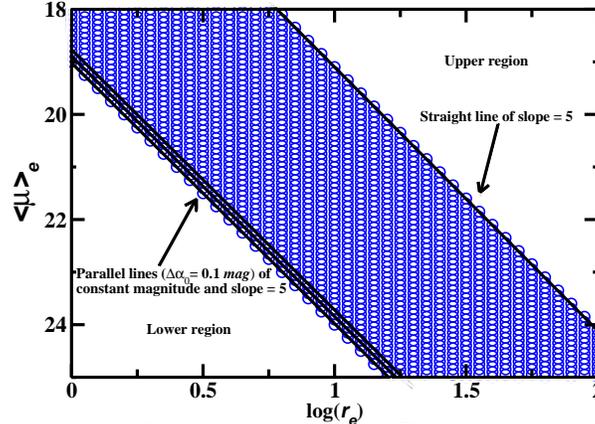}

 \vspace*{-50pt}

  \caption{Distribution of simulated galaxies on the $log(r_{e}) - <\protect\mu>_{e}$ plane. The data distribution consists in parallel lines of slope 5 which shift to brighter magnitude.  }

\end{figure*}


There are several non-parametric methods for the evaluation of sets of data (\cite{ben66}). One of the most popular is the chi-square test. This test measures the discrepancy between an observed probability density and a theoretical probability density (i.e. a normal distribution). Another important test, which does not {\it a priori} assume a specific distribution, is that known as run test. There is one particularly interesting non-parametric test (mean value test), which helps in finding out whether the data in question are distributed at random around a given value or whether this distribution is not a random one. For full details on the hypothesis tests see appendix C. 

In Table 2 we show the results of the application of the tests to the
different galaxy samples in increasing and narrow magnitude intervals. The null hypothesis of the mean value test (test 1) is that $\beta$ has a normal distribution and that its mean value is 5 (we consider the data to have a measurement error equal to 10\%), the null hypothesis of the run test (test 2) is that there is not an underlying trend in the $\beta$ data and finally the null hypothesis of the chi-square test (test 3) is that the $\beta$ data are
random and that they follow a normal distribution. The percentages given in Table 2 refer to the confidence level with which we can reject the null hypothesis.

From Table 2 we can see that, on average, the null hypothesis may be
rejected with a level of confidence of 95 \%. This implies that there are strong reasons to believe that the mean value of $\beta$ is not 5, that there is an underlying trend in the values of $\beta$ and that the distribution of these values is not normal.

From the previous results there is a question that arise: Why is there an underlying trend in the values of $\beta $ when we consider increasing and narrow magnitude intervals? A possible answer to this  
question is that the change in the value of the slope might be due to the fact that the distribution of the galaxies on the log($r_{e}$) - $<\mu>_{e}$ plane depends on the luminosity (Fig. 3, see further details in \cite{var04,don06,nig07}). It could also be due to the geometrical shape of the galaxy distribution on this plane. If the distribution of galaxies takes a rectangular shape, fitting a
straight line to these data will produce different results from a
straight-line fit to a galaxy distribution that takes a
triangular shape or any other shape. In the following section we present simulations of the galaxy distribution on the log($r_{e}$) - $<\mu>_{e}$ plane that elucidate clearly the effects that the shape of the distribution of galaxies on this plane has over the values of the coefficients of the KR.
 
Here, it is important to note that the behaviour of the parameters of the KR in the bright regime is similar for all the samples, in other words, the use of the de Vaucouleurs profiles as an approximation of the S\'ersic profiles (for bright galaxies in some samples) does not affect appreciably the behaviour of the parameters of the KR (see Fig. B.2 and Table B.3). For example, if we compare the data from the brightest part ($M_{B} \lesssim$ -18) of the Abell sample (S\'ersic profile) with the rest of our samples (de Vaucouleurs profile) we
find that the average difference between the $\beta$ coefficients is 9\%, which amount to the size of the errors. On the other hand, when we compare the behaviour of the $\beta$ coefficient for the heterogeneous SDSS sample and the homogeneous SDSS sample (Figs. 1 and 2) we note that this behaviour is similar for both samples
except for the brightest part, however, it is precisely for these galaxies for which the photometry bias could be more pronounced. So it is not possible to say that the differences found are produced
by evolution effects.

\section{Simulations of the galaxy distribution on the log($r_{e}$) - $<\mu>_{e}$ plane}

To investigate whether the dependence of the KR coefficients on the magnitude range is due to the geometric shape of the distribution of galaxies on the log($r_{e}$) - $<\mu>_{e}$ plane (geometrical effect), we perform simulations for each one of our samples of galaxies (Fig. 4). The simulations consist in giving values to log$(r_{e})$ in a similar radii range as that of the sample in
question, fixing the faintest zero-point ($\alpha_{0}$) of this sample (it is equivalent to fixing a magnitude) and fixing a slope of 5 (expected slope considering the definition of total absolute magnitude). Then, we take the same range of log$(r_{e})$, a $\alpha_{0}$ slightly brighter (we take increments of $0.1\;mag$) and the same slope of 5, and so on until we cover the whole range of brightness of the sample in question. The resulting data
distribution consists in parallel lines of slope 5 which shift to brighter magnitude. Once the data are generated, we simulate the observed depopulation effect on the upper region of the galaxy distribution (Fig. 5). This depopulation is known as the exclusion zone (\cite{ben92}) and may be characterized by a straight line (Line of Avoidance or LOA) that has a slope approximately equal to 2.7 (\cite{don06}). Since we do not know the way in which galaxies locations in the lower part of the log($r_{e}$) - $<\mu>_{e}$ plane behave, we consider the following 3 cases in the simulation of this region:

Case 1. It consists in setting a fixed limit-radius for all the magnitude intervals, so that the galaxies are contained within a triangle, one of whose sides is parallel to the $<\mu>_{e}$ axis.

Case 2. In this case we consider that the lower region of the diagram is limited by a 2.7-slope straight line, that is, there is an identical exclusion zone in this part of the diagram, as that observed in the upper part. Therefore the galaxies appear to be contained within a parallelogram and their distribution appears to be symmetric with respect to an axis that contains the barycenter of the galaxy distribution and is parallel to the log$(r_{e})$ axis, we shall call this axis from now on the $X_{Bright}$-axis.
The galaxy distribution is symmetrical under a reflection with respect to the $X_{Bright}$-axis and a $180^{o}$ rotation with respect to a perpendicular line to the $X_{bright}$-axis that contains the barycenter of the galaxy distribution (see Fig. 6). 

Case 3. We consider the lower region as limited in brightness, so galaxies are contained within a triangle with one of its sides parallel to the log$(r_{e})$ axis.


\begin{figure*}

\centering

\includegraphics[bb= 20 20 700 700,angle=-90,width=12cm,clip]{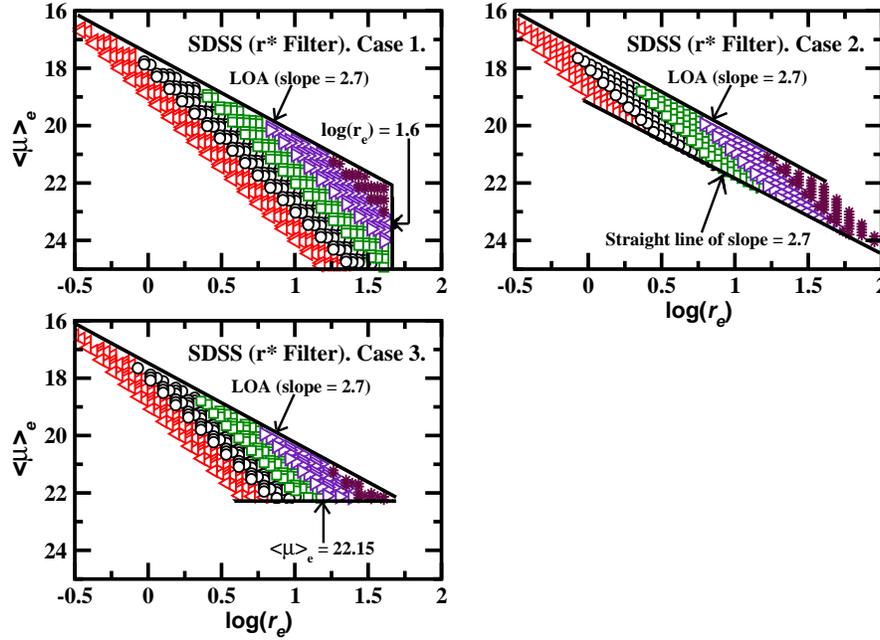}

 \vspace*{-75pt}

  \caption{Distribution of simulated SDSS galaxies (r* filter) on the $log(r_{e}) - <\protect\mu>_{e}$ plane. Each symbol represents a $1$ $mag$ interval. The circles represent the (-21, -22] $mag$
interval. Case 1: upper region of the diagram, limited by a straight line of slope 2.7 (LOA) and radius limit ($log(r_{e}) \le1.6$). Case 2: upper and lower regions of the diagram limited by a straight line of slope 2.7. Case 3: upper region of the diagram limited by a straight line of slope 2.7 and brightness limit ($<\protect\mu>_{e}$ $\le $ $22.15 $ $mag/arcsec^{2}$).}

\end{figure*}



\begin{figure*}

\centering

\includegraphics[bb= 20 20 700 700,angle=-90,width=10cm,clip]{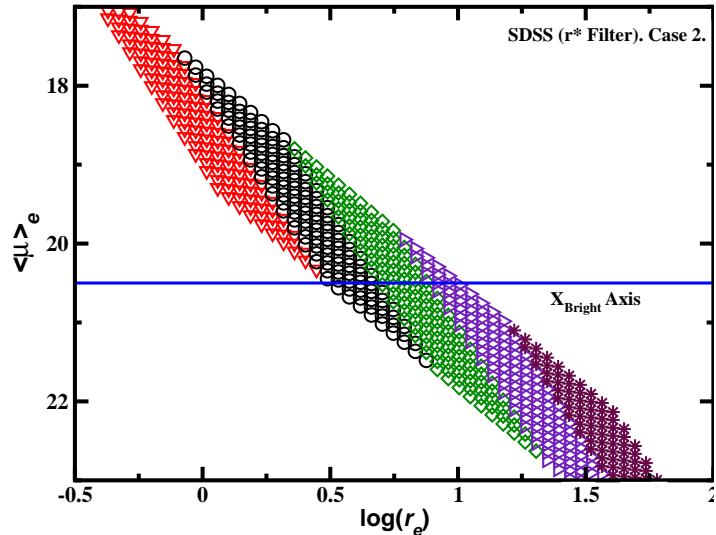}

 \vspace*{-60pt}

  \caption{Distribution of simulated SDSS galaxies (r* filter) on the $log(r_{e}) - <\protect\mu>_{e}$ plane (Case 2). Each symbol represents a $1$ $mag$ interval. The solid line ($X_{Bright}$-axis) is a line of slope=0 that contains the barycenter of the distribution of the galaxies.}

\end{figure*}



\begin{figure*}

\centering

\includegraphics[bb= 20 20 580 500,angle=-90,width=12cm,clip]
{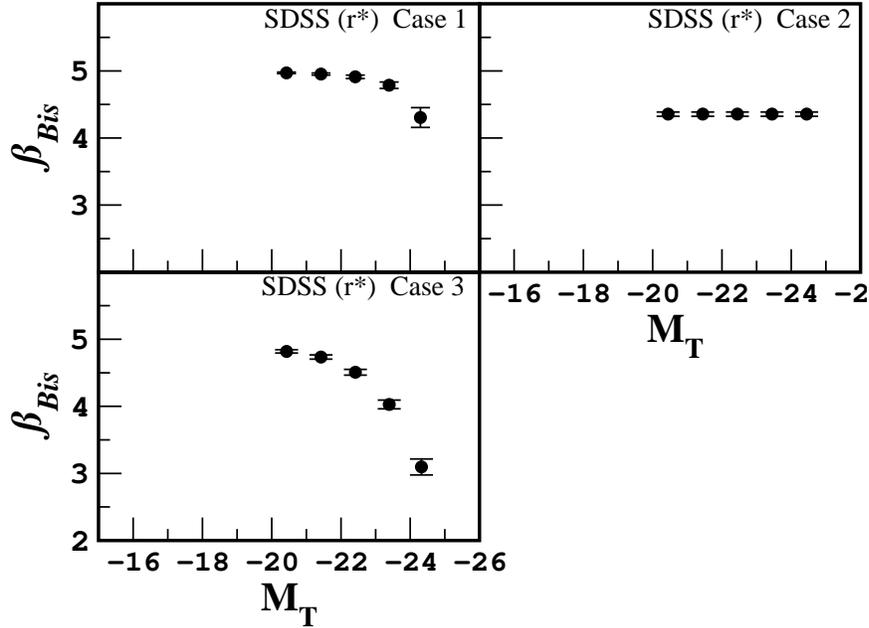}

\vspace*{-10pt}

  \caption{Variation of the KR slope ($\protect\beta$) for one of the SDSS simulations (r* filter). Graphs show the data for the $\protect\beta$ coefficient obtained by the $BCES_{Bis}$ method. Each point represents a $1$ $mag$ interval (mean value of the total absolute magnitude of the galaxies contained in each magnitude range analyzed, see table A.5). }

         \label{FigVibStab}

\end{figure*}


The results of the analysis of the KR show that when we consider increasing magnitude intervals (Tables A.4 and B.4), there are slope changes in all the cases, and these changes are always larger than the errors. We also find that the geometric shape of the distribution of galaxies changes each time we include brighter galaxies in the samples (Fig. 5). On the other hand, when we consider narrow 1-magnitude intervals (Tables A.5 and B.5), there are also slope changes (except for case 2), however, the changes are less pronounced but still larger than the errors and if the magnitude intervals are
progressively narrower, then the changes diminish considerably getting ever closer to the value of 5, just as it occurs for the real samples. Finally, if the magnitude interval is equal to 0.1 $mag$ then the slope is exactly equal to 5 since that is the way we define the samples.

It is important to mention that when we consider narrow magnitude intervals we are able to reproduce, in a reasonable manner, the $\beta$ coefficient variations and the underlying trend found for the real samples (except for case 2) (see Figs. 7 and B.3). We also find that this trend seems to have a maximum around absolute magnitude $M_{B}\sim -18\pm1$. Finally, we find that the geometric shape of the distribution of galaxies changes systematically when we consider brighter magnitude intervals (except for case 2) (Fig. 5). We
must remind the reader that Case 2 corresponds to a symmetrical galaxy distribution over the log($r_{e}$) - $<\mu>_{e}$ plane.

Moreover, we find that both the zero point ($\alpha$) and the intrinsic dispersion ($\sigma_{KR}$) of KR change systematically when we consider brighter galaxies. This latter result confirms the dependence of the intrinsic dispersion with the magnitude range as reported in Nigoche-Netro et al. 2007.

From the aforementioned, we may infer that the KR coefficients and its intrinsic dispersion depend on the width and brightness of the magnitude range. This dependence is caused by a geometrical effect due to the fact that the distribution of the galaxies on the log($r_{e}$) - $<\mu>_{e}$ plane depends on luminosity and also to the fact that this distribution is not symmetrical, in other words, the geometric shape of the distribution of galaxies on the log($r_{e}$) - $<\mu>_{e}$ plane changes systematically as we consider brighter galaxies, and so the values of the KR coefficients change too, because the fitting of a straight line to a set of data does not give the same result for slope and intercept values for data distributed with a
rectangular shape as for data distributed with a triangular shape or, for that matter, with another geometrical shape. In this sense, any other systematic restrictions imposed on a sample of galaxies, such as brightness cuts or effective radius cuts, will cause changes in the geometric shape of the distribution of galaxies. The more pronounced these changes are made, the more pronounced will be the changes in the values of the KR coefficients.

\section{Conclusions}

We have compiled 4 samples of ETGs with information for their photometric parameters log($r_{e}$) and $<\mu>_{e}$, these samples contain a total of $\sim9400$ galaxies in a relatively ample magnitude range ($<\Delta M>$ $\sim 6$ $mag$). From the values of their photometric parameters, we have made an analysis of the behaviour of the coefficients and intrinsic dispersion of the
KR with respect to several characteristics of the magnitude range within which the galaxies are contained. The results from this study are presented as follows:

\begin{itemize}

\item We find that when we include in the samples galaxies which get
progressively brighter (increasing magnitude intervals) or if we consider galaxy samples in progressively brighter fixed-width magnitude intervals (narrow magnitude intervals), the KR coefficients change and these changes result to be larger than the associated errors for most of the cases. We also find that the distribution of the values of the $\beta$ coefficient in narrow magnitude intervals might have a maximum at $M_{B}\sim -18\pm1$. We perform non-parametric tests on the $\beta$ coefficient data and they indicate that the variations are real and that there is evidence of an underlying trend, that is, there is evidence that the $\beta$ coefficient changes systematically when we consider brighter
galaxies.

\item We perform simulations of the different samples of galaxies under study. The results of the analysis of the variation of the KR coefficients, both in increasing and in narrow magnitude intervals show that the coefficients depend on the width and brightness of the magnitude range and that this dependence comes as a result of a geometrical effect due to the fact that:

\begin{itemize}
\item The distribution of galaxies on the log($r_{e}$) - $<\mu>_{e}$ plane depends on luminosity, and

\item That the geometric shape of the distribution of the galaxies on this plane is not symmetrical (see Sect. 4 for full details).

\end{itemize}

\item Finally, simulations confirm the fact that the intrinsic
dispersion of the KR depends on the magnitude range, as asserted in
Nigoche-Netro {\it et al.} (2007).

\end{itemize}

From the previously mentioned results, it is very important to establish that if the magnitude range is not taken into consideration when performing comparisons of galaxy samples such as the dependence of the KR on the environment, on redshift or on wavelength, the differences which might be found may be misinterpreted.

\section*{Acknowledgments}

We would like to thank Project WINGS (\cite{fas02,var04}) for
providing the photometric parameters for the samples in 7 Abell clusters, Consejo Nacional de Ciencia y Tecnolog\'{\i}a (M\'{e}xico) for a PhD fellowship number 132526, Ministerio Espa\~{n}ol de Educaci\'{o}n y Ciencia for grant PNAYA2006, Instituto de Matem\'{a}ticas y F\'{\i}sica Fundamental (CSIC, Espa\~{n}a) and Instituto de Astronom\'{\i}a (UNAM, M\'{e}xico) for all the facilities provided for the realization of this project. We would also like to record our obligation to Prof. Mariano Moles for help with the presentation of this paper. Last but not least, we would like to express our deepest appreciation to an anonymous referee whose comments and suggestions greatly improved the presentation of this paper.

\begin{appendix}

\section{Tables with behaviour of the KR coefficients with respect to absolute magnitude range. The information presented here corresponds to the original photometry of the different galaxy samples.}



\begin{table*}

{\tiny

\begin{minipage}{185mm}

\caption{KR coefficients for the different galaxy samples in increasing intervals of magnitude (upper magnitude cut-off).
MI is the total absolute magnitude interval within which the galaxies are distributed, N is the number of galaxies in the magnitude interval, $\alpha_{Bis}$ is the zero point of KR, $\beta_{Bis}$ is the slope of KR, $\sigma_{KR}$ is the intrinsic dispersion of KR and R is the correlation coefficient of the fit (Pearson Statistics). }

  \begin{tabular}{@{}cccccccccccccc@{}}
\hline

 MI &  N   &  $\alpha_{Bis}$      &  $\beta_{Bis}$  &  $\sigma_{KR}$ &  R && MI &  N   &  $\alpha_{Bis}$      &  $\beta_{Bis}$  &  $\sigma_{KR}$ &  R \\

\hline

&&&&&&&&&&&&\\

&&Abell (V)&&&&&&&Coma (Gunn r)&&\\

\cline{1-6} \cline{8-13} \\

-16 $\geq M_{V}$  \textgreater -17 & 136    &21.997$\pm$0.063&4.733$\pm$0.268& 0.201  & 0.933 && -17 $\geq M_{Gr}$ \textgreater -18 & 17      &20.019$\pm$0.138&2.537$\pm$0.276&0.536&0.802 \\
-16 $\geq M_{V}$ \textgreater -18 & 284    &21.232$\pm$0.075&5.949$\pm$0.274& 0.488  & 0.834 && -17 $\geq M_{Gr}$ \textgreater -19 & 68     &19.377$\pm$0.111&3.266$\pm$0.313&0.692&0.705 \\
-16 $\geq M_{V}$ \textgreater -19 & 394   &20.917$\pm$0.116&5.721$\pm$0.411& 0.719  & 0.715  && -17 $\geq M_{Gr}$ \textgreater -20 & 82     &19.302$\pm$0.095&3.281$\pm$0.229 &0.667&0.769 \\
-16 $\geq M_{V}$  \textgreater -20 & 519    &20.574$\pm$0.190&5.216$\pm$0.630& 1.031  & 0.504 && -17 $\geq M_{Gr}$ \textgreater -21 & 109     &19.135$\pm$0.093&3.210$\pm$0.219& 0.656 & 0.743  \\
-16 $\geq M_{V}$  \textgreater -21 & 593   &20.967$\pm$0.216&3.102$\pm$0.673& 1.143  & 0.326 && -17 $\geq M_{Gr}$ \textgreater -22 & 169     &18.917$\pm$0.090&2.982$\pm$0.167& 0.674 & 0.725  \\
-16 $\geq M_{V}$  \textgreater -22 & 626    &21.229$\pm$0.155&2.070$\pm$0.445& 1.143  & 0.251 && -17 $\geq M_{Gr}$ \textgreater -23 & 192      &18.824$\pm$0.092&2.687$\pm$0.173& 0.711 & 0.645  \\
... & ...  & ... & ... & ...  & ...  && -17 $\geq M_{Gr}$  \textgreater -24 & 196      &18.837$\pm$0.089&2.598$\pm$0.158& 0.708 & 0.651  \\

&&&&&&&&&&&&\\

&&Hydra (Gunn r)&&&&&&&&&\\

\cline{1-6} \\

-18 $\geq M_{Gr}$  \textgreater -19 & 4     &19.722$\pm$0.024&5.037$\pm$0.062& 0.077  & 0.996 \\
-18 $\geq M_{Gr}$  \textgreater -20 & 15    &18.964$\pm$0.122&5.721$\pm$0.158& 0.378  & 0.961 \\
-18 $\geq M_{Gr}$  \textgreater -21 & 42    &18.295$\pm$0.151&5.277$\pm$0.213& 0.685  & 0.885 \\
-18 $\geq M_{Gr}$  \textgreater -22 & 54    &18.068$\pm$0.170&5.097$\pm$0.256& 0.824  & 0.813 \\

&&&&&&&&&&&&\\

&&SDSS (g*)&&&&&&&SDSS (r*)&&\\

\cline{1-6} \cline{8-13} \\

 -18.5 $\geq M_{g*}$  \textgreater -19 & 26    &18.392$\pm$0.040&4.808$\pm$0.100& 0.128  & 0.982 && -18.5 $\geq M_{r*}$  \textgreater -19 & 3    &...&...&...&... \\
 -18.5 $\geq M_{g*}$  \textgreater -20 & 517    &18.978$\pm$0.027&4.726$\pm$0.057& 0.297  & 0.900 && -18.5 $\geq M_{r*}$  \textgreater -20 & 77      &18.962$\pm$0.051&4.645$\pm$0.138&0.258&0.929 \\
 -18.5 $\geq M_{g*}$  \textgreater -21 & 2856   &18.686$\pm$0.020&3.909$\pm$0.033& 0.362  & 0.825 && -18.5 $\geq M_{r*}$  \textgreater -21 & 1014     &18.204$\pm$0.025&4.356$\pm$0.050&0.322&0.877 \\
 -18.5 $\geq M_{g*}$  \textgreater -22 & 6687    &18.790$\pm$0.016&3.057$\pm$0.021& 0.406  & 0.763 && -18.5 $\geq M_{r*}$  \textgreater -22 & 4057     &17.963$\pm$0.019&3.609$\pm$0.030 &0.386&0.784 \\
 -18.5 $\geq M_{g*}$  \textgreater -23 & 8516    &18.949$\pm$0.014&2.641$\pm$0.016& 0.425  & 0.744 && -18.5 $\geq M_{r*}$  \textgreater -23 & 7646     &18.093$\pm$0.015&2.858$\pm$0.020& 0.418 & 0.724  \\
 -18.5 $\geq M_{g*}$  \textgreater -24 & 8661    &18.989$\pm$0.013&2.572$\pm$0.015& 0.425  & 0.749 && -18.5 $\geq M_{r*}$  \textgreater -24 & 8630     &18.214$\pm$0.013&2.572$\pm$0.017& 0.426 & 0.717  \\
 ... &  ...  & ...& ... & ...  & ... && -18.5 $\geq M_{r*}$  \textgreater -24.7 & 8666      &18.225$\pm$0.013&2.552$\pm$0.016& 0.426 & 0.721  \\

&&&&&&&&&&&&\\

&&SDSS (i*)&&&&&&&SDSS (z*)&&\\

\cline{1-6} \cline{8-13} \\

 -19 $\geq M_{i*}$  \textgreater -20 & 31    &17.458$\pm$0.059&4.717$\pm$0.209& 0.219  & 0.937 && -19.5 $\geq M_{z*}$  \textgreater -20 & 15    &17.331$\pm$0.038&4.482$\pm$0.104 &0.116 &0.984 \\
 -19 $\geq M_{i*}$  \textgreater -21 & 505    &18.033$\pm$0.027&4.540$\pm$0.062& 0.311  & 0.894 && -19.5 $\geq M_{z*}$  \textgreater -21 & 210      &17.940$\pm$0.035&4.742$\pm$0.085&0.299&0.908 \\
 -19 $\geq M_{i*}$  \textgreater -22 & 2805   &17.605$\pm$0.020&3.941$\pm$0.036& 0.371  & 0.816 && -19.5 $\geq M_{z*}$  \textgreater -22 & 1799    &17.312$\pm$0.022&4.259$\pm$0.042&0.358&0.850 \\
 -19 $\geq M_{i*}$  \textgreater -23 & 6645    &17.663$\pm$0.015&3.047$\pm$0.022& 0.406  & 0.741 && -19.5 $\geq M_{z*}$  \textgreater -23 & 5440    &17.264$\pm$0.017&3.309$\pm$0.025 &0.400&0.765 \\
 -19 $\geq M_{i*}$  \textgreater -24 & 8512    &17.808$\pm$0.013&2.598$\pm$0.013& 0.422  & 0.716 && -19.5 $\geq M_{z*}$  \textgreater -24 & 8277   &17.437$\pm$0.013&2.680$\pm$0.017& 0.414 & 0.738  \\
 -19 $\geq M_{i*}$  \textgreater -25 & 8664    &17.843$\pm$0.012&2.529$\pm$0.016& 0.422  & 0.723 && -19.5 $\geq M_{z*}$  \textgreater -25 & 8658   &17.504$\pm$0.012&2.543$\pm$0.015& 0.413 & 0.748  \\
 ... &  ...  & ...& ... & ...  & ... && -19.5 $\geq M_{z*}$  \textgreater -25.3 & 8665 &17.507$\pm$0.012&2.537$\pm$0.015& 0.413 & 0.750  \\

&&&&&&&&&&&&\\

\hline

\end{tabular}
\end{minipage}

}

\end{table*}





\begin{table*}

{\tiny


\begin{minipage}{185mm}

\caption{KR coefficients for the different galaxy samples in increasing intervals of magnitude (lower magnitude cut-off).
MI is the total absolute magnitude interval within which the galaxies are distributed, N is the number of galaxies in the magnitude interval, $\alpha_{Bis}$ is the zero point of KR, $\beta_{Bis}$ is the slope of KR, $\sigma_{KR}$ is the intrinsic dispersion of KR and R is the correlation coefficient of the fit (Pearson Statistics). }

  \begin{tabular}{@{}cccccccccccccc@{}}
\hline

 MI &  N   &  $\alpha_{Bis}$      &  $\beta_{Bis}$  &  $\sigma_{KR}$ &  R && MI &  N   &  $\alpha_{Bis}$      &  $\beta_{Bis}$  &  $\sigma_{KR}$ &  R \\

\hline

&&&&&&&&&&&&\\

&&Abell (V)&&&&&&&Coma (Gunn r)&&\\

\cline{1-6} \cline{8-13} \\

-21 $\geq M_{V}$  \textgreater -22 & 33    &18.246$\pm$0.071&3.830$\pm$0.072& 0.146  & 0.972 && -23 $\geq M_{Gr}$ \textgreater -24 & 4     &16.813$\pm$0.234&3.569$\pm$0.211&0.243&0.894 \\
-20 $\geq M_{V}$ \textgreater -22 & 107    &18.797$\pm$0.052&3.594$\pm$0.081& 0.355  & 0.902 && -22 $\geq M_{Gr}$ \textgreater -24 & 36      &17.251$\pm$0.126&3.592$\pm$0.188&0.285&0.912 \\
-19 $\geq M_{V}$ \textgreater -22 & 232   &19.371$\pm$0.054&3.346$\pm$0.098& 0.501  & 0.791  && -21 $\geq M_{Gr}$ \textgreater -24 & 87     &17.492$\pm$0.107&3.937$\pm$0.193&0.483&0.829 \\
-18 $\geq M_{V}$  \textgreater -22 & 342    &19.757$\pm$0.111&3.464$\pm$0.270& 0.756  & 0.599 && -20 $\geq M_{Gr}$ \textgreater -24 & 114     &17.900$\pm$0.090&3.562$\pm$0.162 &0.521&0.786 \\
-17 $\geq M_{V}$  \textgreater -22 & 490   &20.551$\pm$0.147&2.751$\pm$0.392& 0.947  & 0.373 && -19 $\geq M_{Gr}$ \textgreater -24 & 128     &18.197$\pm$0.073&3.195$\pm$0.119& 0.550 & 0.796  \\
-16 $\geq M_{V}$  \textgreater -22 & 626    &21.229$\pm$0.155&2.070$\pm$0.445& 1.143  & 0.251 && -18 $\geq M_{Gr}$ \textgreater -24 & 179     &18.628$\pm$0.080&2.832$\pm$0.137& 0.654 & 0.713  \\
... & ...  & ... & ... & ...  & ...  && -17 $\geq M_{Gr}$  \textgreater -24 & 196      &18.837$\pm$0.089&2.598$\pm$0.158& 0.708 & 0.651  \\

&&&&&&&&&&&&\\

&&Hydra (Gunn r)&&&&&&&&&\\

\cline{1-6} \\

-21 $\geq M_{Gr}$  \textgreater -22 & 12     &17.142$\pm$0.112&4.821$\pm$0.148& 0.240  & 0.974 \\
-20 $\geq M_{Gr}$  \textgreater -22 & 39    &17.669$\pm$0.095&5.019$\pm$0.144& 0.490  & 0.934 \\
-19 $\geq M_{Gr}$  \textgreater -22 & 50    &18.092$\pm$0.157&4.803$\pm$0.228& 0.705  & 0.854 \\
-18 $\geq M_{Gr}$  \textgreater -22 & 54    &18.068$\pm$0.170&5.097$\pm$0.256& 0.824  & 0.813 \\

&&&&&&&&&&&&\\

&&SDSS (g*)&&&&&&&SDSS (r*)&&\\

\cline{1-6} \cline{8-13} \\

 -23 $\geq M_{g*}$  \textgreater -24 & 145    &14.993$\pm$0.086&4.118$\pm$0.072& 0.161  & 0.951 && -24 $\geq M_{r*}$  \textgreater -24.7 & 36      &14.081$\pm$0.162&4.162$\pm$0.132&0.151&0.953 \\
 -22 $\geq M_{g*}$  \textgreater -24 & 1974    &17.245$\pm$0.028&3.867$\pm$0.028& 0.261  & 0.902 && -23 $\geq M_{r*}$  \textgreater -24.7 & 1020     &16.272$\pm$0.033&3.954$\pm$0.034&0.217&0.926 \\
 -21 $\geq M_{g*}$  \textgreater -24 & 5805   &18.197$\pm$0.016&3.272$\pm$0.019& 0.347  & 0.826 && -22 $\geq M_{r*}$  \textgreater -24.7 & 4609     &17.180$\pm$0.017&3.477$\pm$0.021 &0.317&0.837 \\
 -20 $\geq M_{g*}$  \textgreater -24 & 8144    &18.793$\pm$0.012&2.765$\pm$0.015& 0.400  & 0.776 && -21 $\geq M_{r*}$  \textgreater -24.7 & 7652     &17.896$\pm$0.012&2.880$\pm$0.015& 0.381 & 0.775  \\
 -19 $\geq M_{g*}$  \textgreater -24 & 8635    &18.972$\pm$0.012&2.589$\pm$0.015& 0.421  & 0.754 && -20 $\geq M_{r*}$  \textgreater -24.7 & 8589     &18.178$\pm$0.012&2.605$\pm$0.015& 0.418 & 0.733  \\
 -18.5 $\geq M_{g*}$  \textgreater -24 & 8661    &18.989$\pm$0.013&2.572$\pm$0.015& 0.425  & 0.749 && -19 $\geq M_{r*}$  \textgreater -24.7 & 8663     &18.222$\pm$0.013&2.557$\pm$0.016& 0.426 & 0.721  \\
 ... &  ...  & ...& ... & ...  & ... && -18.5 $\geq M_{r*}$  \textgreater -24.7 & 8666      &18.225$\pm$0.013&2.552$\pm$0.016& 0.426 & 0.721  \\

&&&&&&&&&&&&\\

&&SDSS (i*)&&&&&&&SDSS (z*)&&\\

\cline{1-6} \cline{8-13} \\

 -24 $\geq M_{i*}$  \textgreater -25 & 152    &13.819$\pm$0.068&4.240$\pm$0.062& 0.160  & 0.956 && -24 $\geq M_{z*}$  \textgreater -25 & 381    &13.915$\pm$0.035&4.027$\pm$0.033 &0.158 &0.953 \\
 -23 $\geq M_{i*}$  \textgreater -25 & 2019    &16.245$\pm$0.024&3.791$\pm$0.027& 0.252  & 0.896 && -23 $\geq M_{z*}$  \textgreater -25 & 3218      &16.274$\pm$0.017&3.562$\pm$0.020&0.264&0.883 \\
 -22 $\geq M_{i*}$  \textgreater -25 & 5859   &17.136$\pm$0.014&3.194$\pm$0.018& 0.337  & 0.811 && -22$\geq M_{z*}$  \textgreater -25 & 6859    &17.038$\pm$0.011&2.985$\pm$0.014&0.340&0.821 \\
 -21 $\geq M_{i*}$  \textgreater -25 & 8159    &17.661$\pm$0.011&2.721$\pm$0.015& 0.394  & 0.757 && -21 $\geq M_{z*}$  \textgreater -25 & 8448    &17.410$\pm$0.011&2.643$\pm$0.013 &0.395&0.771 \\
 -20 $\geq M_{i*}$  \textgreater -25 & 8633    &17.821$\pm$0.012&2.555$\pm$0.015& 0.418  & 0.729 && -20 $\geq M_{z*}$  \textgreater -25 & 8643   &17.493$\pm$0.011&2.555$\pm$0.014& 0.410 & 0.753  \\
 -19 $\geq M_{i*}$  \textgreater -25 & 8664    &17.843$\pm$0.012&2.529$\pm$0.016& 0.422  & 0.723 && -19.5 $\geq M_{z*}$  \textgreater -25 & 8658   &17.504$\pm$0.012&2.543$\pm$0.015& 0.413 & 0.748  \\

&&&&&&&&&&&&\\

\hline

\end{tabular}
\end{minipage}

}

\end{table*}




\begin{table*}

{\tiny

 \begin{minipage}{185mm}
\caption{KR coefficients for the different galaxy samples in narrow $1$ $mag$ intervals. MI is the total absolute magnitude interval within which the galaxies are distributed, N is the number of galaxies in the magnitude interval, $\alpha_{Bis}$ is the zero point of KR, $\beta_{Bis}$ is the slope of KR, $\sigma_{KR}$ is the intrinsic dispersion of KR and R is the correlation coefficient
of the fit (Pearson Statistics). }

  \begin{tabular}{@{}cccccccccccccc@{}}
\hline

 MI &  N   &  $\alpha_{Bis}$      &  $\beta_{Bis}$  &  $\sigma_{KR}$ &  R && MI &  N   &  $\alpha_{Bis}$      &  $\beta_{Bis}$  &  $\sigma_{KR}$ &  R \\

\hline

&&&&&&&&&&&&\\

&&Abell (V)&&&&&&&Coma (Gunn r)&&\\

\cline{1-6} \cline{8-13} \\

 -16.0 $\geq M_{V}$  \textgreater -17.0 & 136    &21.997$\pm$0.063&4.733$\pm$0.268& 0.201  & 0.933 &&-17.0 $\geq M_{Gr}$  \textgreater -18.0 &  17      &20.019$\pm$0.138&2.537$\pm$0.276&0.536&0.802 \\
 -17.0 $\geq M_{V}$  \textgreater -18.0 & 148    &21.009$\pm$0.051&5.276$\pm$0.193& 0.275  & 0.924 &&-18.0 $\geq M_{Gr}$  \textgreater -19.0 &  51     &19.087$\pm$0.111&3.783$\pm$0.256&0.623&0.772 \\
 -18.0 $\geq M_{V}$  \textgreater -19.0 & 110   &20.223$\pm$0.082&4.886$\pm$0.319& 0.310  & 0.918 &&-19.0 $\geq M_{Gr}$  \textgreater -20.0 &  14     &18.957$\pm$0.081&3.243$\pm$0.090 &0.316&0.970 \\
 -19.0 $\geq M_{V}$  \textgreater -20.0 & 125    &19.128$\pm$0.044&4.983$\pm$0.093& 0.287  & 0.929 &&-20.0 $\geq M_{Gr}$  \textgreater -21.0 &  27     &18.015$\pm$0.044&4.839$\pm$0.077& 0.182 & 0.971  \\
 -20.0 $\geq M_{V}$  \textgreater -21.0 & 74    &18.310$\pm$0.070&4.791$\pm$0.120& 0.289  & 0.926 &&-21.0 $\geq M_{Gr}$  \textgreater -22.0 &  51     &17.427$\pm$0.052&4.685$\pm$0.067& 0.251 & 0.965  \\
 -21.0 $\geq M_{V}$  \textgreater -22.0 & 33    &18.246$\pm$0.071&3.830$\pm$0.072& 0.146  & 0.941 &&-22.0 $\geq M_{Gr}$  \textgreater -23.0 & 32      &16.879$\pm$0.035&4.512$\pm$0.156& 0.221 & 0.936  \\
 ... & ...  & ...& ... & ...  & ...    &&-23.0 $\geq M_{Gr}$  \textgreater -24.0 & 4      &17.087$\pm$0.057&3.569$\pm$0.211& 0.243 & 0.894  \\

&&&&&&&&&&&&\\

&&Hydra (Gunn r)&&&&&&&&&\\

\cline{1-6} \\

 -18.0 $\geq M_{Gr}$  \textgreater -19.0 & 4     &19.722$\pm$0.024&5.037$\pm$0.062& 0.077  & 0.996 \\
 -19.0 $\geq M_{Gr}$  \textgreater -20.0 & 11    &18.994$\pm$0.115&5.255$\pm$0.167& 0.312  & 0.968 \\
 -20.0 $\geq M_{Gr}$  \textgreater -21.0 & 27    &17.902$\pm$0.063&5.127$\pm$0.078& 0.237  & 0.987 \\
 -21.0 $\geq M_{Gr}$  \textgreater -22.0 & 12    &17.142$\pm$0.112&4.821$\pm$0.148& 0.240  & 0.974 \\

&&&&&&&&&&&&\\

&&SDSS (g*)&&&&&&&SDSS (r*)&&\\

\cline{1-6} \cline{8-13} \\

 -18.5 $\geq M_{g*}$  \textgreater -19.0 & 26 &18.392$\pm$0.040&4.808$\pm$0.100& 0.128 & 0.982 &&-18.5 $\geq M_{r*}$  \textgreater -19.0 &   3  & ...&  ... &  ...  & ... \\
 -19.0 $\geq M_{g*}$  \textgreater -20.0 & 491 &18.922$\pm$0.021&4.760$\pm$0.044& 0.238 & 0.936 &&-19.0 $\geq M_{r*}$  \textgreater -20.0 &  74 &18.864$\pm$0.035&4.966$\pm$0.091& 0.220 & 0.948  \\
 -20.0 $\geq M_{g*}$  \textgreater -21.0 & 2339&18.254$\pm$0.013&4.463$\pm$0.021& 0.249 & 0.918 &&-20.0 $\geq M_{r*}$  \textgreater -21.0 &  937&18.036$\pm$0.014&4.619$\pm$0.028& 0.233 & 0.937  \\
 -21.0 $\geq M_{g*}$  \textgreater -22.0 & 3831&17.590$\pm$0.014&4.277$\pm$0.017& 0.245 & 0.912 &&-21.0 $\geq M_{r*}$  \textgreater -22.0 &  3043 &17.366$\pm$0.011&4.342$\pm$0.018& 0.244 & 0.914  \\
 -22.0 $\geq M_{g*}$  \textgreater -23.0 & 1829&16.874$\pm$0.023&4.279$\pm$0.023& 0.223 & 0.927 &&-22.0 $\geq M_{r*}$  \textgreater -23.0 &  3589&16.624$\pm$0.014&4.320$\pm$0.017& 0.242 & 0.905  \\
 -23.0 $\geq M_{g*}$  \textgreater -24.0 & 145 &16.386$\pm$0.086& 4.118$\pm$0.072  & 0.161  & 0.951 &&-23.0 $\geq M_{r*}$  \textgreater -24.0 &  984 &16.016$\pm$0.027&4.224$\pm$0.028& 0.199 & 0.934  \\
 ... &  ... & ...&  ... &  ...  &  ...  &&-24.0 $\geq M_{r*}$  \textgreater -24.7 &  36&15.474$\pm$0.162&4.162$\pm$0.132& 0.151 & 0.953  \\

&&&&&&&&&&&&\\

&&SDSS (i*)&&&&&&&SDSS (z*)&&\\

\cline{1-6} \cline{8-13} \\

 -19.0 $\geq M_{i*}$  \textgreater -20.0 & 31      &17.458$\pm$0.059& 4.717$\pm$0.209  &    0.219 & 0.937   &&-19.5 $\geq M_{z*}$  \textgreater -20.0 & 15  &17.331$\pm$0.038&4.482$\pm$0.104 & 0.116 & 0.984  \\   
 -20.0 $\geq M_{i*}$  \textgreater -21.0 & 474     &17.926$\pm$0.018& 4.713$\pm$0.039  &    0.235 & 0.941   &&-20.0 $\geq M_{z*}$  \textgreater -21.0 &   195  &17.859$\pm$0.024&4.838$\pm$0.056 & 0.229 & 0.986  \\
 -21.0 $\geq M_{i*}$  \textgreater -22.0 & 2300     &17.213$\pm$0.012& 4.481$\pm$0.021  &    0.249 & 0.918   &&-21.0 $\geq M_{z*}$  \textgreater -22.0 &  1589  &17.065$\pm$0.013&4.610$\pm$0.024 & 0.246 & 0.980  \\
 -22.0 $\geq M_{i*}$  \textgreater -23.0 & 3840     &16.572$\pm$0.012& 4.250$\pm$0.016  &    0.241 & 0.904   &&-22.0 $\geq M_{z*}$  \textgreater -23.0 &  3641  &16.475$\pm$0.011&4.213$\pm$0.015 & 0.236 & 0.980  \\
 -23.0 $\geq M_{i*}$  \textgreater -24.0 & 1867    &15.877$\pm$0.020& 4.239$\pm$0.023  &    0.219 & 0.918   &&-23.0 $\geq M_{z*}$  \textgreater -24.0 &  2837 &15.831$\pm$0.015&4.142$\pm$0.018 & 0.223 & 0.977  \\
 -24.0 $\geq M_{i*}$  \textgreater -25.0 & 152 &15.212$\pm$0.068& 4.240$\pm$0.062      & 0.160 & 0.956 &&-24.0 $\geq M_{z*}$  \textgreater -25.0 &  381 &15.308$\pm$0.035&4.027$\pm$0.033 & 0.158 & 0.973  \\

&&&&&&&&&&&&\\

\hline

\end{tabular}
\end{minipage}

}

\end{table*}


\clearpage


\begin{table*}
 \centering

{\tiny

 \begin{minipage}{175mm}
\caption{KR coefficients for the SDSS simulation in r* filter. Increasing magnitude intervals. MI is the total absolute magnitude interval within which the galaxies are distributed, N is the number of galaxies in the magnitude interval, $\alpha_{Bis}$ is the zero point of KR, $\beta_{Bis}$ is the slope of KR, $\sigma_{KR}$ is the intrinsic dispersion of KR and R is the correlation coefficient
of the fit (Pearson Statistics).}

  \begin{tabular}{@{}cccccccccccccc@{}}
\hline

 MI &  N   &  $\alpha_{Bis}$      &  $\beta_{Bis}$  &  $\sigma_{KR}$ &  R && MI &  N   &  $\alpha_{Bis}$      &  $\beta_{Bis}$  &  $\sigma_{KR}$ &  R \\

\hline

&&&&&&&&&&&&\\

&&Case 1&&&&&&&Case 2&&\\

\cline{1-6} \cline{8-13} \\

-20 $\geq M_{r*}$  \textgreater -21 & 455    &18.587$\pm$0.015&4.971$\pm$0.011& 0.285  & 0.995 &&-20 $\geq M_{r*}$  \textgreater -21 &  140    &18.533$\pm$0.020&4.356$\pm$0.032 &0.234 &0.968 \\
-20 $\geq M_{r*}$  \textgreater -22 & 810    &18.232$\pm$0.022&4.865$\pm$0.017& 0.555  & 0.977 &&-20 $\geq M_{r*}$  \textgreater -22 &  280     &18.313$\pm$0.020&3.604$\pm$0.023&0.330&0.953 \\
-20 $\geq M_{r*}$  \textgreater -23 & 1065   &18.029$\pm$0.025&4.673$\pm$0.022& 0.785  & 0.945 &&-20 $\geq M_{r*}$  \textgreater -23 &  420    &18.270$\pm$0.019&3.213$\pm$0.018&0.365&0.961\\
-20 $\geq M_{r*}$  \textgreater -24 & 1220   &17.977$\pm$0.025&4.442$\pm$0.027& 0.953  & 0.907 && -20 $\geq M_{r*}$  \textgreater -24 & 560    &18.277$\pm$0.018&3.014$\pm$0.013 &0.381&0.971 \\
-20 $\geq M_{r*}$  \textgreater -25 & 1275  &17.995$\pm$0.025&4.305$\pm$0.031& 1.030  & 0.885 && -20 $\geq M_{r*}$  \textgreater -25 & 676   &18.311$\pm$0.017&2.872$\pm$0.011& 0.388 & 0.976  \\

&&&&&&&&&&&&\\

&&Case 3&&&&&&&&&\\

\cline{1-6} \\

-20 $\geq M_{r*}$  \textgreater -21 &   245 &18.602$\pm$0.018& 4.819$\pm$0.025 &0.275 &0.984\\
-20 $\geq M_{r*}$  \textgreater -22 & 439     &18.360$\pm$0.022&4.305$\pm$0.027& 0.478  & 0.943 \\
-20 $\geq M_{r*}$  \textgreater -23 & 579   &18.338$\pm$0.023&3.708$\pm$0.028& 0.585  & 0.906 \\
-20 $\geq M_{r*}$  \textgreater -24 & 664   &18.412$\pm$0.026&3.268$\pm$0.029& 0.627  & 0.889 \\
-20 $\geq M_{r*}$  \textgreater -25 & 669   &18.474$\pm$0.027&3.053$\pm$0.028& 0.640  & 0.887 \\

&&&&&&&&&&&&\\

\hline

\end{tabular}
\end{minipage}

}

\end{table*}


\begin{table*}
 \centering

{\tiny

 \begin{minipage}{175mm}
\caption{KR coefficients for the SDSS simulation for r* filter. Narrow $1$ $mag$ intervals. MI is the total absolute magnitude interval within which the galaxies are distributed, N is the number of galaxies in the magnitude interval, $\alpha_{Bis}$ is the zero point of KR, $\beta_{Bis}$ is the slope of KR, $\sigma_{KR}$ is the intrinsic dispersion of KR and R is the correlation coefficient of the fit (Pearson Statistics).}

  \begin{tabular}{@{}cccccccccccccc@{}}
\hline

 MI &  N   &  $\alpha_{Bis}$      &  $\beta_{Bis}$  &  $\sigma_{KR}$ &  R && MI &  N   &  $\alpha_{Bis}$      &  $\beta_{Bis}$  &  $\sigma_{KR}$ &  R \\

\hline

&&&&&&&&&&&&\\

&&Case 1&&&&&&&Case 2&&\\

\cline{1-6} \cline{8-13} \\

 -20 $\geq M_{r*}$  \textgreater -21 & 455    &18.587$\pm$0.015&4.971$\pm$0.011& 0.285  & 0.995 &&-20 $\geq M_{r*}$  \textgreater -21 & 140    &18.533$\pm$0.020&4.356$\pm$0.032& 0.234 & 0.968   \\
 -21 $\geq M_{r*}$  \textgreater -22 & 355    &17.614$\pm$0.020&4.953$\pm$0.016& 0.284  & 0.992 &&-21 $\geq M_{r*}$  \textgreater -22 & 140     &17.810$\pm$0.024&4.356$\pm$0.032& 0.234 & 0.968  \\
 -22 $\geq M_{r*}$  \textgreater -23 & 255   &16.677$\pm$0.031&4.912$\pm$0.025& 0.280 & 0.985 &&-22 $\geq M_{r*}$  \textgreater -23 & 140    &17.087$\pm$0.033&4.356$\pm$0.032& 0.234 & 0.968  \\
 -23 $\geq M_{r*}$  \textgreater -24 & 155   &15.878$\pm$0.060&4.786$\pm$0.048& 0.270  & 0.963 &&-23 $\geq M_{r*}$  \textgreater -24 & 140    &16.364$\pm$0.045&4.356$\pm$0.032& 0.234 & 0.968  \\
 -24 $\geq M_{r*}$  \textgreater -25 & 55  &15.727$\pm$0.200&4.305$\pm$0.147& 0.212  & 0.885 &&-24 $\geq M_{r*}$  \textgreater -25 & 140  &15.788$\pm$0.069&4.356$\pm$0.032& 0.234 & 0.968  \\

&&&&&&&&&&&&\\

&&Case 3&&&&&&&&&\\

\cline{1-6} \\

 -20 $\geq M_{r*}$  \textgreater -21 &   245 &18.602$\pm$0.018& 4.819$\pm$0.025 &0.275 &0.984\\
 -21 $\geq M_{r*}$  \textgreater -22 & 194    &17.709$\pm$0.024&4.735$\pm$0.032& 0.269  & 0.976 \\
 -22 $\geq M_{r*}$  \textgreater -23 & 140   &16.984$\pm$0.037&4.508$\pm$0.043& 0.253  & 0.960 \\
 -23 $\geq M_{r*}$  \textgreater -24 & 85   &16.697$\pm$0.068&4.028$\pm$0.065& 0.218 & 0.925\\
 -24 $\geq M_{r*}$  \textgreater -25 & 35   &17.391$\pm$0.168&3.097$\pm$0.119& 0.140  & 0.891 \\

&&&&&&&&&&&&\\

\hline

\end{tabular}
\end{minipage}

}

\end{table*}


\clearpage

\end{appendix}


\begin{appendix}

\section{Figures and Tables with behaviour of the KR coefficients with respect to absolute magnitude range. The information presented here corresponds to photometry in the B filter.}

\clearpage


\begin{figure*}

\centering

\includegraphics[bb= 20 20 750 750,angle=-90,width=12cm,clip]{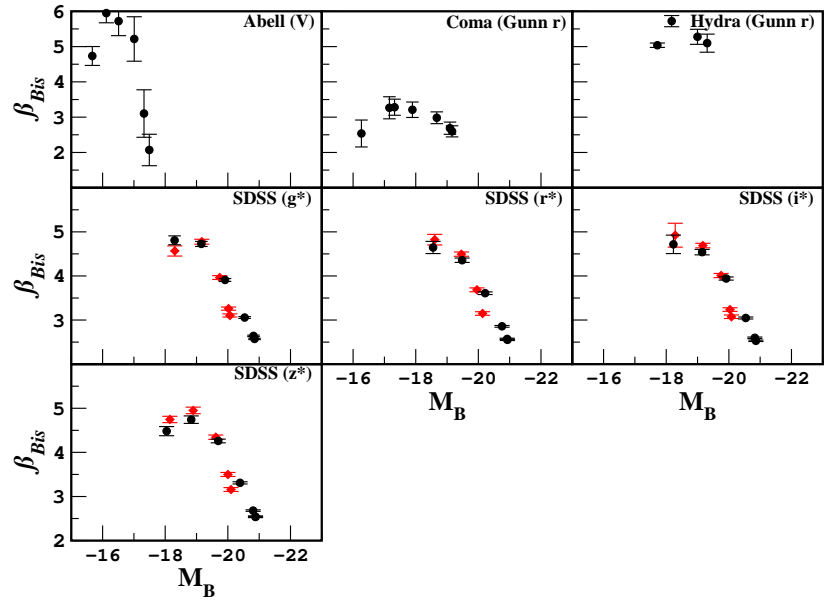}

 \vspace*{-90pt}

  \caption{Equivalent to Fig. 1 with photometric information in the B-filter. Variation of the KR slope ($\protect\beta$) for the different samples of galaxies in increasing magnitude intervals (see Table B.1). Diamonds represent the SDSS homogeneous sample.}

\end{figure*}



\begin{figure*}

\centering

\includegraphics[bb= 20 20 750 750,angle=-90,width=12cm,clip]{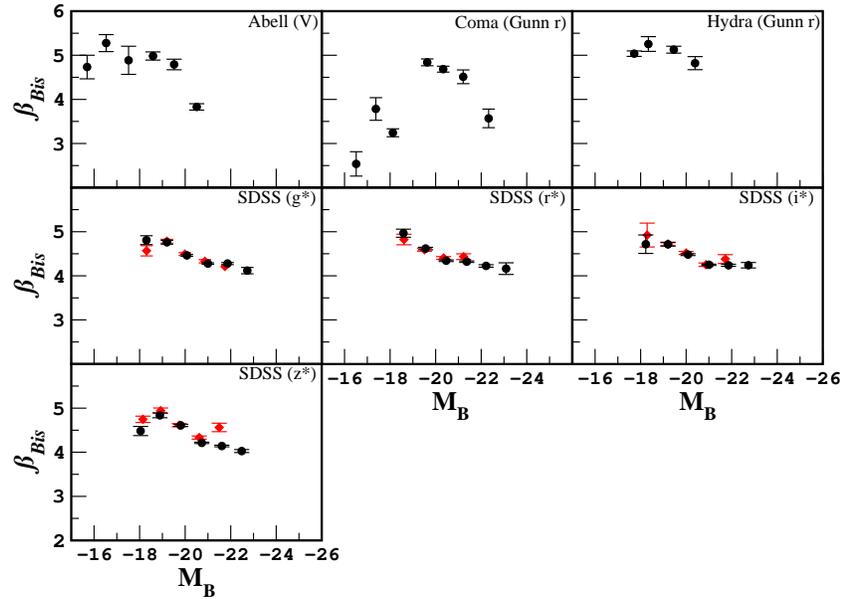}

 \vspace*{-90pt}

  \caption{Equivalent to Fig. 2 with photometric information in the B-filter. Variation of the KR slope ($\protect\beta$) for the different samples of galaxies. Each point represents a $1$ $mag$ interval (see Table B.3). Diamonds represent the SDSS homogeneous sample.}

\end{figure*}



\begin{figure*}

\centering

\includegraphics[bb= 20 20 580 500,angle=-90,width=12cm,clip]{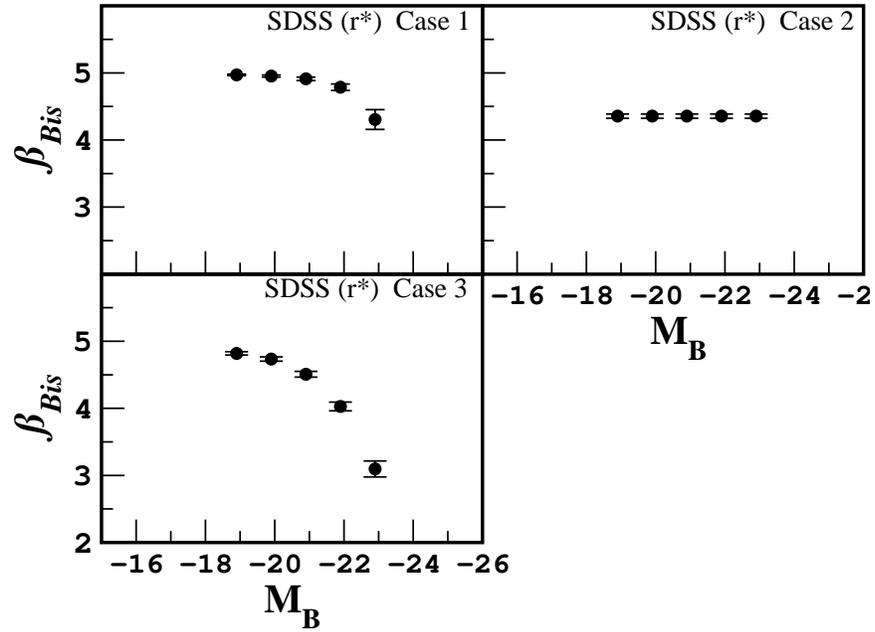}

\vspace*{-10pt}

  \caption{Equivalent to Fig. 7 with photometric information in the B-filter. Variation of the KR slope ($\protect\beta$) for one of the SDSS simulations. Each point represents a $1$ $mag$ interval (see Table B.5).}

         \label{FigVibStab}

\end{figure*}





\begin{table*}

{\tiny

\begin{minipage}{185mm}

\caption{Equivalent to Table A.1 with photometric information in the B-filter. KR coefficients for the different galaxy samples in increasing intervals of magnitude (upper magnitude cut-off). MI is the total absolute magnitude interval within which the galaxies are distributed, N is the number of galaxies in the magnitude interval, $\alpha_{Bis}$ is the zero point of KR, $\beta_{Bis}$ is the slope of KR, $\sigma_{KR}$ is the intrinsic dispersion of KR and R is the correlation coefficient of the fit (Pearson Statistics). }

  \begin{tabular}{@{}cccccccccccccc@{}}
\hline

 MI &  N   &  $\alpha_{Bis}$      &  $\beta_{Bis}$  &  $\sigma_{KR}$ &  R && MI &  N   &  $\alpha_{Bis}$      &  $\beta_{Bis}$  &  $\sigma_{KR}$ &  R \\

\hline

&&&&&&&&&&&&\\

&&Abell (V)&&&&&&&Coma (Gunn r)&&\\

\cline{1-6} \cline{8-13} \\

-15.1 $\geq M_{B}$  \textgreater -16.1 & 136    &22.917$\pm$0.063&4.733$\pm$0.268& 0.201  & 0.933 && -15.9 $\geq M_{B}$ \textgreater -16.9 & 17      &21.169$\pm$0.138&2.537$\pm$0.276&0.536&0.802 \\
-15.1 $\geq M_{B}$ \textgreater -17.1 & 284    &22.152$\pm$0.075&5.949$\pm$0.274& 0.488  & 0.834 && -15.9 $\geq M_{B}$ \textgreater -17.9 & 68     &20.527$\pm$0.111&3.266$\pm$0.313&0.692&0.705 \\
-15.1 $\geq M_{B}$ \textgreater -18.1 & 394   &21.837$\pm$0.116&5.721$\pm$0.411& 0.719  & 0.715  && -15.9 $\geq M_{B}$ \textgreater -18.9 & 82     &20.452$\pm$0.095&3.281$\pm$0.229 &0.667&0.769 \\
-15.1 $\geq M_{B}$  \textgreater -19.1 & 519    &21.494$\pm$0.190&5.216$\pm$0.630& 1.031  & 0.504 && -15.9 $\geq M_{B}$ \textgreater -19.9 & 109     &20.285$\pm$0.093&3.210$\pm$0.219& 0.656 & 0.743  \\
-15.1 $\geq M_{B}$  \textgreater -20.1 & 593   &21.887$\pm$0.216&3.102$\pm$0.673& 1.143  & 0.326 && -15.9 $\geq M_{B}$ \textgreater -20.9 & 169     &20.067$\pm$0.090&2.982$\pm$0.167& 0.674 & 0.725  \\
-15.1 $\geq M_{B}$  \textgreater -21.1 & 626    &22.149$\pm$0.155&2.070$\pm$0.445& 1.143  & 0.251 && -15.9 $\geq M_{B}$ \textgreater -21.9 & 192      &19.974$\pm$0.092&2.687$\pm$0.173& 0.711 & 0.645  \\
... & ...  & ... & ... & ...  & ...  && -15.9 $\geq M_{B}$  \textgreater -22.9 & 196      &19.987$\pm$0.089&2.598$\pm$0.158& 0.708 & 0.651  \\

&&&&&&&&&&&&\\

&&Hydra (Gunn r)&&&&&&&&&\\

\cline{1-6} \\

-16.9 $\geq M_{B}$  \textgreater -17.9 & 4     &20.872$\pm$0.024&5.037$\pm$0.062& 0.077  & 0.996 \\
-16.9 $\geq M_{B}$  \textgreater -18.9 & 15    &20.114$\pm$0.122&5.721$\pm$0.158& 0.378  & 0.961 \\
-16.9 $\geq M_{B}$  \textgreater -19.9 & 42    &19.445$\pm$0.151&5.277$\pm$0.213& 0.685  & 0.885 \\
-16.9 $\geq M_{B}$  \textgreater -20.9 & 54    &19.218$\pm$0.170&5.097$\pm$0.256& 0.824  & 0.813 \\

&&&&&&&&&&&&\\

&&SDSS (g*)&&&&&&&SDSS (r*)&&\\

\cline{1-6} \cline{8-13} \\

 -18.0 $\geq M_{B}$  \textgreater -18.5 & 26    &18.892$\pm$0.040&4.808$\pm$0.100& 0.128  & 0.982 && -17.4 $\geq M_{B}$  \textgreater -17.9 & 3    &...&...&...&... \\
 -18.0 $\geq M_{B}$  \textgreater -19.5 & 517    &19.478$\pm$0.027&4.726$\pm$0.057& 0.297  & 0.900 && -17.4 $\geq M_{B}$  \textgreater -18.9 & 77      &20.062$\pm$0.051&4.645$\pm$0.138&0.258&0.929 \\
 -18.0 $\geq M_{B}$  \textgreater -20.5 & 2856   &19.186$\pm$0.020&3.909$\pm$0.033& 0.362  & 0.825 && -17.4 $\geq M_{B}$  \textgreater -19.9 & 1014     &19.304$\pm$0.025&4.356$\pm$0.050&0.322&0.877 \\
 -18.0 $\geq M_{B}$  \textgreater -21.5 & 6687    &19.290$\pm$0.016&3.057$\pm$0.021& 0.406  & 0.763 && -17.4 $\geq M_{B}$  \textgreater -20.9 & 4057     &19.063$\pm$0.019&3.609$\pm$0.030 &0.386&0.784 \\
 -18.0 $\geq M_{B}$  \textgreater -22.5 & 8516    &19.449$\pm$0.014&2.641$\pm$0.016& 0.425  & 0.744 && -17.4 $\geq M_{B}$  \textgreater -21.9 & 7646     &19.193$\pm$0.015&2.858$\pm$0.020& 0.418 & 0.724  \\
 -18.0 $\geq M_{B}$  \textgreater -23.5 & 8661    &19.489$\pm$0.013&2.572$\pm$0.015& 0.425  & 0.749 && -17.4 $\geq M_{B}$  \textgreater -22.9 & 8630     &19.314$\pm$0.013&2.572$\pm$0.017& 0.426 & 0.717  \\
 ... &  ...  & ...& ... & ...  & ... && -17.4 $\geq M_{B}$  \textgreater -23.6 & 8666      &19.325$\pm$0.013&2.552$\pm$0.016& 0.426 & 0.721  \\

&&&&&&&&&&&&\\

&&SDSS (i*)&&&&&&&SDSS (z*)&&\\

\cline{1-6} \cline{8-13} \\

 -17.5 $\geq M_{B}$  \textgreater -18.5 & 31    &18.958$\pm$0.059&4.717$\pm$0.209& 0.219  & 0.937 && -17.7 $\geq M_{B}$  \textgreater -18.2 & 15    &19.131$\pm$0.038&4.482$\pm$0.104 &0.116 &0.984 \\
 -17.5 $\geq M_{B}$  \textgreater -19.5 & 505    &19.533$\pm$0.027&4.540$\pm$0.062& 0.311  & 0.894 && -17.7 $\geq M_{B}$  \textgreater -19.2 & 210      &19.740$\pm$0.035&4.742$\pm$0.085&0.299&0.908 \\
 -17.5 $\geq M_{B}$  \textgreater -20.5 & 2805   &19.105$\pm$0.020&3.941$\pm$0.036& 0.371  & 0.816 && -17.7 $\geq M_{B}$  \textgreater -20.2 & 1799    &19.112$\pm$0.022&4.259$\pm$0.042&0.358&0.850 \\
 -17.5 $\geq M_{B}$  \textgreater -21.5 & 6645    &19.163$\pm$0.015&3.047$\pm$0.022& 0.406  & 0.741 && -17.7 $\geq M_{B}$  \textgreater -21.2 & 5440    &19.064$\pm$0.017&3.309$\pm$0.025 &0.400&0.765 \\
 -17.5 $\geq M_{B}$  \textgreater -22.5 & 8512    &19.308$\pm$0.013&2.598$\pm$0.013& 0.422  & 0.716 && -17.7 $\geq M_{B}$  \textgreater -22.2 & 8277   &19.237$\pm$0.013&2.680$\pm$0.017& 0.414 & 0.738  \\
 -17.5 $\geq M_{B}$  \textgreater -23.5 & 8664    &19.343$\pm$0.012&2.529$\pm$0.016& 0.422  & 0.723 && -17.7 $\geq M_{B}$  \textgreater -23.2 & 8658   &19.304$\pm$0.012&2.543$\pm$0.015& 0.413 & 0.748  \\
 ... &  ...  & ...& ... & ...  & ... && -17.7 $\geq M_{B}$  \textgreater -23.5 & 8665 &19.307$\pm$0.012&2.537$\pm$0.015& 0.413 & 0.750  \\

&&&&&&&&&&&&\\

\hline

\end{tabular}
\end{minipage}

}

\end{table*}





\begin{table*}

{\tiny


\begin{minipage}{185mm}

\caption{Equivalent to Table A.2 with photometric information in the B-filter. KR coefficients for the different galaxy samples in increasing intervals of magnitude (lower magnitude cut-off). MI is the total absolute magnitude interval within which the galaxies are distributed, N is the number of galaxies in the magnitude interval, $\alpha_{Bis}$ is the zero point of KR, $\beta_{Bis}$ is the slope of KR, $\sigma_{KR}$ is the intrinsic dispersion of KR and R is the correlation coefficient of the fit (Pearson Statistics). }

  \begin{tabular}{@{}cccccccccccccc@{}}
\hline

 MI &  N   &  $\alpha_{Bis}$      &  $\beta_{Bis}$  &  $\sigma_{KR}$ &  R && MI &  N   &  $\alpha_{Bis}$      &  $\beta_{Bis}$  &  $\sigma_{KR}$ &  R \\

\hline

&&&&&&&&&&&&\\

&&Abell (V)&&&&&&&Coma (Gunn r)&&\\

\cline{1-6} \cline{8-13} \\

-20.1 $\geq M_{B}$  \textgreater -21.1 & 33    &19.166$\pm$0.071&3.830$\pm$0.072& 0.146  & 0.972 && -21.9 $\geq M_{B}$ \textgreater -22.9 & 4     &17.963$\pm$0.234&3.569$\pm$0.211&0.243&0.894 \\
-19.1 $\geq M_{B}$ \textgreater -21.1 & 107    &19.717$\pm$0.052&3.594$\pm$0.081& 0.355  & 0.902 && -20.9 $\geq M_{B}$ \textgreater -22.9 & 36      &18.401$\pm$0.126&3.592$\pm$0.188&0.285&0.912 \\
-18.1 $\geq M_{B}$ \textgreater -21.1 & 232   &20.291$\pm$0.054&3.346$\pm$0.098& 0.501  & 0.791  && -19.9 $\geq M_{B}$ \textgreater -22.9 & 87     &18.642$\pm$0.107&3.937$\pm$0.193&0.483&0.829 \\
-17.1 $\geq M_{B}$  \textgreater -21.1 & 342    &20.677$\pm$0.111&3.464$\pm$0.270& 0.756  & 0.599 && -18.9 $\geq M_{B}$ \textgreater -22.9 & 114     &19.050$\pm$0.090&3.562$\pm$0.162 &0.521&0.786 \\
-16.1 $\geq M_{B}$  \textgreater -21.1 & 490   &21.471$\pm$0.147&2.751$\pm$0.392& 0.947  & 0.373 && -17.9 $\geq M_{B}$ \textgreater -22.9 & 128     &19.347$\pm$0.073&3.195$\pm$0.119& 0.550 & 0.796  \\
-15.1 $\geq M_{B}$  \textgreater -21.1 & 626    &22.149$\pm$0.155&2.070$\pm$0.445& 1.143  & 0.251 && -16.9 $\geq M_{B}$ \textgreater -22.9 & 179     &19.778$\pm$0.080&2.832$\pm$0.137& 0.654 & 0.713  \\
... & ...  & ... & ... & ...  & ...  && -15.9 $\geq M_{B}$  \textgreater -22.9 & 196      &19.987$\pm$0.089&2.598$\pm$0.158& 0.708 & 0.651  \\

&&&&&&&&&&&&\\

&&Hydra (Gunn r)&&&&&&&&&\\

\cline{1-6} \\

-19.9 $\geq M_{B}$  \textgreater -20.9 & 12     &18.292$\pm$0.112&4.821$\pm$0.148& 0.240  & 0.974 \\
-18.9 $\geq M_{B}$  \textgreater -20.9 & 39    &18.819$\pm$0.095&5.019$\pm$0.144& 0.490  & 0.934 \\
-17.9 $\geq M_{B}$  \textgreater -20.9 & 50    &19.242$\pm$0.157&4.803$\pm$0.228& 0.705  & 0.854 \\
-16.9 $\geq M_{B}$  \textgreater -20.9 & 54    &19.218$\pm$0.170&5.097$\pm$0.256& 0.824  & 0.813 \\

&&&&&&&&&&&&\\

&&SDSS (g*)&&&&&&&SDSS (r*)&&\\

\cline{1-6} \cline{8-13} \\

 -22.5 $\geq M_{B}$  \textgreater -23.5 & 145    &15.493$\pm$0.086&4.118$\pm$0.072& 0.161  & 0.951 && -22.9 $\geq M_{B}$  \textgreater -23.6 & 36      &15.181$\pm$0.162&4.162$\pm$0.132&0.151&0.953 \\
 -21.5 $\geq M_{B}$  \textgreater -23.5 & 1974    &17.745$\pm$0.028&3.867$\pm$0.028& 0.261  & 0.902 && -21.9 $\geq M_{B}$  \textgreater -23.6 & 1020     &17.372$\pm$0.033&3.954$\pm$0.034&0.217&0.926 \\
 -20.5 $\geq M_{B}$  \textgreater -23.5 & 5805   &18.697$\pm$0.016&3.272$\pm$0.019& 0.347  & 0.826 && -20.9 $\geq M_{B}$  \textgreater -23.6 & 4609     &18.280$\pm$0.017&3.477$\pm$0.021 &0.317&0.837 \\
 -19.5 $\geq M_{B}$  \textgreater -23.5 & 8144    &19.293$\pm$0.012&2.765$\pm$0.015& 0.400  & 0.776 && -19.9 $\geq M_{B}$  \textgreater -23.6 & 7652     &18.996$\pm$0.012&2.880$\pm$0.015& 0.381 & 0.775  \\
 -18.5 $\geq M_{B}$  \textgreater -23.5 & 8635    &19.472$\pm$0.012&2.589$\pm$0.015& 0.421  & 0.754 && -18.9 $\geq M_{B}$  \textgreater -23.6 & 8589     &19.278$\pm$0.012&2.605$\pm$0.015& 0.418 & 0.733  \\
 -18.0 $\geq M_{B}$  \textgreater -23.5 & 8661    &19.489$\pm$0.013&2.572$\pm$0.015& 0.425  & 0.749 && -17.9 $\geq M_{B}$  \textgreater -23.6 & 8663     &19.322$\pm$0.013&2.557$\pm$0.016& 0.426 & 0.721  \\
 ... &  ...  & ...& ... & ...  & ... && -17.4 $\geq M_{B}$  \textgreater -23.6 & 8666      &19.325$\pm$0.013&2.552$\pm$0.016& 0.426 & 0.721  \\

&&&&&&&&&&&&\\

&&SDSS (i*)&&&&&&&SDSS (z*)&&\\

\cline{1-6} \cline{8-13} \\

 -22.5 $\geq M_{B}$  \textgreater -23.5 & 152    &15.319$\pm$0.068&4.240$\pm$0.062& 0.160  & 0.956 && -22.2 $\geq M_{B}$  \textgreater -23.2 & 381    &15.715$\pm$0.035&4.027$\pm$0.033 &0.158 &0.953 \\
 -21.5 $\geq M_{B}$  \textgreater -23.5 & 2019    &17.745$\pm$0.024&3.791$\pm$0.027& 0.252  & 0.896 && -21.2 $\geq M_{B}$  \textgreater -23.2 & 3218      &18.074$\pm$0.017&3.562$\pm$0.020&0.264&0.883 \\
 -20.5 $\geq M_{B}$  \textgreater -23.5 & 5859   &18.636$\pm$0.014&3.194$\pm$0.018& 0.337  & 0.811 && -20.2 $\geq M_{B}$  \textgreater -23.2 & 6859    &18.838$\pm$0.011&2.985$\pm$0.014&0.340&0.821 \\
 -19.5 $\geq M_{B}$  \textgreater -23.5 & 8159    &19.161$\pm$0.011&2.721$\pm$0.015& 0.394  & 0.757 && -19.2 $\geq M_{B}$  \textgreater -23.2 & 8448    &19.210$\pm$0.011&2.643$\pm$0.013 &0.395&0.771 \\
 -18.5 $\geq M_{B}$  \textgreater -23.5 & 8633    &19.321$\pm$0.012&2.555$\pm$0.015& 0.418  & 0.729 && -18.2 $\geq M_{B}$  \textgreater -23.2 & 8643   &19.293$\pm$0.011&2.555$\pm$0.014& 0.410 & 0.753  \\
 -17.5 $\geq M_{B}$  \textgreater -23.5 & 8664    &19.343$\pm$0.012&2.529$\pm$0.016& 0.422  & 0.723 && -17.7 $\geq M_{B}$  \textgreater -23.2 & 8658   &19.304$\pm$0.012&2.543$\pm$0.015& 0.413 & 0.748  \\

&&&&&&&&&&&&\\

\hline

\end{tabular}
\end{minipage}

}

\end{table*}




\begin{table*}

{\tiny

 \begin{minipage}{185mm}
\caption{Equivalent to Table A.3 with photometric information in the B-filter. KR coefficients for the different galaxy samples in narrow $1$ $mag$ intervals. MI is the total absolute magnitude interval within which the galaxies are distributed, N is the number of galaxies in the magnitude interval, $\alpha_{Bis}$ is the zero point of KR, $\beta_{Bis}$ is the slope of KR, $\sigma_{KR}$ is the intrinsic dispersion of KR and R is the correlation coefficient
of the fit (Pearson Statistics). }

  \begin{tabular}{@{}cccccccccccccc@{}}
\hline

 MI &  N   &  $\alpha_{Bis}$      &  $\beta_{Bis}$  &  $\sigma_{KR}$ &  R && MI &  N   &  $\alpha_{Bis}$      &  $\beta_{Bis}$  &  $\sigma_{KR}$ &  R \\

\hline

&&&&&&&&&&&&\\

&&Abell (V)&&&&&&&Coma (Gunn r)&&\\

\cline{1-6} \cline{8-13} \\

 -15.1 $\geq M_{B}$  \textgreater -16.1 & 136    &22.917$\pm$0.063&4.733$\pm$0.268& 0.201  & 0.933 &&-15.9 $\geq M_{B}$  \textgreater -16.9 &  17      &21.169$\pm$0.138&2.537$\pm$0.276&0.536&0.802 \\
 -16.1 $\geq M_{B}$  \textgreater -17.1 & 148    &21.929$\pm$0.051&5.276$\pm$0.193& 0.275  & 0.924 &&-16.9 $\geq M_{B}$  \textgreater -17.9 &  51     &20.237$\pm$0.111&3.783$\pm$0.256&0.623&0.772 \\
 -17.1 $\geq M_{B}$  \textgreater -18.1 & 110   &21.143$\pm$0.082&4.886$\pm$0.319& 0.310  & 0.918 &&-17.9 $\geq M_{B}$  \textgreater -18.9 &  14     &20.107$\pm$0.081&3.243$\pm$0.090 &0.316&0.970 \\
 -18.1 $\geq M_{B}$  \textgreater -19.1 & 125    &20.048$\pm$0.044&4.983$\pm$0.093& 0.287  & 0.929 &&-18.9 $\geq M_{B}$  \textgreater -19.9 &  27     &19.165$\pm$0.044&4.839$\pm$0.077& 0.182 & 0.971  \\
 -19.1 $\geq M_{B}$  \textgreater -20.1 & 74    &19.230$\pm$0.070&4.791$\pm$0.120& 0.289  & 0.926 &&-19.9 $\geq M_{B}$  \textgreater -20.9 &  51     &18.577$\pm$0.052&4.685$\pm$0.067& 0.251 & 0.965  \\
 -20.1 $\geq M_{B}$  \textgreater -21.1 & 33    &19.166$\pm$0.071&3.830$\pm$0.072& 0.146  & 0.941 &&-20.9 $\geq M_{B}$  \textgreater -21.9 & 32      &18.029$\pm$0.035&4.512$\pm$0.156& 0.221 & 0.936  \\
 ... & ...  & ...& ... & ...  & ...    &&-21.9 $\geq M_{B}$  \textgreater -22.9 & 4      &18.237$\pm$0.057&3.569$\pm$0.211& 0.243 & 0.894  \\

&&&&&&&&&&&&\\

&&Hydra (Gunn r)&&&&&&&&&\\

\cline{1-6} \\

 -16.9 $\geq M_{B}$  \textgreater -17.9 & 4     &20.872$\pm$0.024&5.037$\pm$0.062& 0.077  & 0.996 \\
 -17.9 $\geq M_{B}$  \textgreater -18.9 & 11    &20.144$\pm$0.115&5.255$\pm$0.167& 0.312  & 0.968 \\
 -18.9 $\geq M_{B}$  \textgreater -19.9 & 27    &19.052$\pm$0.063&5.127$\pm$0.078& 0.237  & 0.987 \\
 -19.9 $\geq M_{B}$  \textgreater -20.9 & 12    &18.292$\pm$0.112&4.821$\pm$0.148& 0.240  & 0.974 \\

&&&&&&&&&&&&\\

&&SDSS (g*)&&&&&&&SDSS (r*)&&\\

\cline{1-6} \cline{8-13} \\

 -18.0 $\geq M_{B}$  \textgreater -18.5 & 26 &18.892$\pm$0.040&4.808$\pm$0.100& 0.128 & 0.982 &&-17.4 $\geq M_{B}$  \textgreater -17.9 &   3  & ...&  ... &  ...  & ... \\
 -18.5 $\geq M_{B}$  \textgreater -19.5 & 491 &19.492$\pm$0.021&4.760$\pm$0.044& 0.238 & 0.936 &&-17.9 $\geq M_{B}$  \textgreater -18.9 &  74 &19.964$\pm$0.035&4.966$\pm$0.091& 0.220 & 0.948  \\
 -19.5 $\geq M_{B}$  \textgreater -20.5 & 2339&18.754$\pm$0.013&4.463$\pm$0.021& 0.249 & 0.918 &&-18.9 $\geq M_{B}$  \textgreater -19.9 &  937&19.136$\pm$0.014&4.619$\pm$0.028& 0.233 & 0.937  \\
 -20.5 $\geq M_{B}$  \textgreater -21.5 & 3831&18.090$\pm$0.014&4.277$\pm$0.017& 0.245 & 0.912 &&-19.9 $\geq M_{B}$  \textgreater -20.9 &  3043 &18.466$\pm$0.011&4.342$\pm$0.018& 0.244 & 0.914  \\
 -21.5 $\geq M_{B}$  \textgreater -22.5 & 1829&17.374$\pm$0.023&4.279$\pm$0.023& 0.223 & 0.927 &&-20.9 $\geq M_{B}$  \textgreater -21.9 &  3589&17.724$\pm$0.014&4.320$\pm$0.017& 0.242 & 0.905  \\
 -22.5 $\geq M_{B}$  \textgreater -23.5 & 145 &16.886$\pm$0.086& 4.118$\pm$0.072  & 0.161  & 0.951 &&-21.9 $\geq M_{B}$  \textgreater -22.9 &  984 &17.116$\pm$0.027&4.224$\pm$0.028& 0.199 & 0.934  \\
 ... &  ... & ...&  ... &  ...  &  ...  &&-22.9 $\geq M_{B}$  \textgreater -23.6 &  36&16.574$\pm$0.162&4.162$\pm$0.132& 0.151 & 0.953  \\

&&&&&&&&&&&&\\

&&SDSS (i*)&&&&&&&SDSS (z*)&&\\

\cline{1-6} \cline{8-13} \\

 -17.5 $\geq M_{B}$  \textgreater -18.5 & 31      &18.958$\pm$0.059& 4.717$\pm$0.209  &    0.219 & 0.937   &&-17.7 $\geq M_{B}$  \textgreater -18.2 & 15  &19.131$\pm$0.038&4.482$\pm$0.104 & 0.116 & 0.984  \\   
 -18.5 $\geq M_{B}$  \textgreater -19.5 & 474     &19.426$\pm$0.018& 4.713$\pm$0.039  &    0.235 & 0.941   &&-18.2 $\geq M_{B}$  \textgreater -19.2 &   195  &19.659$\pm$0.024&4.838$\pm$0.056 & 0.229 & 0.986  \\
 -19.5 $\geq M_{B}$  \textgreater -20.5 & 2300     &18.713$\pm$0.012& 4.481$\pm$0.021  &    0.249 & 0.918   &&-19.2 $\geq M_{B}$  \textgreater -20.2 &  1589  &18.865$\pm$0.013&4.610$\pm$0.024 & 0.246 & 0.980  \\
 -20.5 $\geq M_{B}$  \textgreater -21.5 & 3840     &18.072$\pm$0.012& 4.250$\pm$0.016  &    0.241 & 0.904   &&-20.2 $\geq M_{B}$  \textgreater -21.2 &  3641  &18.275$\pm$0.011&4.213$\pm$0.015 & 0.236 & 0.980  \\
 -21.5 $\geq M_{B}$  \textgreater -22.5 & 1867    &17.377$\pm$0.020& 4.239$\pm$0.023  &    0.219 & 0.918   &&-21.2 $\geq M_{B}$  \textgreater -22.2 &  2837 &17.631$\pm$0.015&4.142$\pm$0.018 & 0.223 & 0.977  \\
 -22.5 $\geq M_{B}$  \textgreater -23.5 & 152 &16.712$\pm$0.068& 4.240$\pm$0.062      & 0.160 & 0.956 &&-22.2 $\geq M_{B}$  \textgreater -23.2 &  381 &17.108$\pm$0.035&4.027$\pm$0.033 & 0.158 & 0.973  \\

&&&&&&&&&&&&\\

\hline

\end{tabular}
\end{minipage}

}

\end{table*}


\clearpage


\begin{table*}
 \centering

{\tiny

 \begin{minipage}{175mm}
\caption{Equivalent to Table A.4 with photometric information in the B-filter. KR coefficients for the SDSS simulation in r* filter. Increasing magnitude intervals. MI is the total absolute magnitude interval within which the galaxies are distributed, N is the number of galaxies in the magnitude interval, $\alpha_{Bis}$ is the zero point of KR, $\beta_{Bis}$ is the slope of KR, $\sigma_{KR}$ is the intrinsic dispersion of KR and R is the correlation coefficient
of the fit (Pearson Statistics).}

  \begin{tabular}{@{}cccccccccccccc@{}}
\hline

 MI &  N   &  $\alpha_{Bis}$      &  $\beta_{Bis}$  &  $\sigma_{KR}$ &  R && MI &  N   &  $\alpha_{Bis}$      &  $\beta_{Bis}$  &  $\sigma_{KR}$ &  R \\

\hline

&&&&&&&&&&&&\\

&&Case 1&&&&&&&Case 2&&\\

\cline{1-6} \cline{8-13} \\

-18.9 $\geq M_{B}$  \textgreater -19.9 & 455    &19.687$\pm$0.015&4.971$\pm$0.011& 0.285  & 0.995 &&-18.9 $\geq M_{B}$  \textgreater -19.9 &  140    &19.633$\pm$0.020&4.356$\pm$0.032 &0.234 &0.968 \\
-18.9 $\geq M_{B}$  \textgreater -20.9 & 810    &19.332$\pm$0.022&4.865$\pm$0.017& 0.555  & 0.977 &&-18.9 $\geq M_{B}$  \textgreater -20.9 &  280     &19.413$\pm$0.020&3.604$\pm$0.023&0.330&0.953 \\
-18.9 $\geq M_{B}$  \textgreater -21.9 & 1065   &19.129$\pm$0.025&4.673$\pm$0.022& 0.785  & 0.945 &&-18.9 $\geq M_{B}$  \textgreater -21.9 &  420    &19.370$\pm$0.019&3.213$\pm$0.018&0.365&0.961\\
-18.9 $\geq M_{B}$  \textgreater -22.9 & 1220   &19.077$\pm$0.025&4.442$\pm$0.027& 0.953  & 0.907 && -18.9 $\geq M_{B}$  \textgreater -22.9 & 560    &19.377$\pm$0.018&3.014$\pm$0.013 &0.381&0.971 \\
-18.9 $\geq M_{B}$  \textgreater -23.9 & 1275  &19.095$\pm$0.025&4.305$\pm$0.031& 1.030  & 0.885 && -18.9 $\geq M_{B}$  \textgreater -23.9 & 676   &19.411$\pm$0.017&2.872$\pm$0.011& 0.388 & 0.976  \\

&&&&&&&&&&&&\\

&&Case 3&&&&&&&&&\\

\cline{1-6} \\

-18.9 $\geq M_{B}$  \textgreater -19.9 &   245 &19.702$\pm$0.018& 4.819$\pm$0.025 &0.275 &0.984\\
-18.9 $\geq M_{B}$  \textgreater -20.9 & 439     &19.460$\pm$0.022&4.305$\pm$0.027& 0.478  & 0.943 \\
-18.9 $\geq M_{B}$  \textgreater -21.9 & 579   &19.438$\pm$0.023&3.708$\pm$0.028& 0.585  & 0.906 \\
-18.9 $\geq M_{B}$  \textgreater -22.9 & 664   &19.512$\pm$0.026&3.268$\pm$0.029& 0.627  & 0.889 \\
-18.9 $\geq M_{B}$  \textgreater -23.9 & 669   &19.574$\pm$0.027&3.053$\pm$0.028& 0.640  & 0.887 \\

&&&&&&&&&&&&\\

\hline

\end{tabular}
\end{minipage}

}

\end{table*}


\begin{table*}
 \centering

{\tiny

 \begin{minipage}{175mm}
\caption{Equivalent to Table A.5 with photometric information in the B-filter. KR coefficients for the SDSS simulation for r* filter. Narrow $1$ $mag$ intervals. MI is the total absolute magnitude interval within which the galaxies are distributed, N is the number of galaxies in the magnitude interval, $\alpha_{Bis}$ is the zero point of KR, $\beta_{Bis}$ is the slope of KR, $\sigma_{KR}$ is the intrinsic dispersion of KR and R is the correlation coefficient of the fit (Pearson Statistics).}

  \begin{tabular}{@{}cccccccccccccc@{}}
\hline

 MI &  N   &  $\alpha_{Bis}$      &  $\beta_{Bis}$  &  $\sigma_{KR}$ &  R && MI &  N   &  $\alpha_{Bis}$      &  $\beta_{Bis}$  &  $\sigma_{KR}$ &  R \\

\hline

&&&&&&&&&&&&\\

&&Case 1&&&&&&&Case 2&&\\

\cline{1-6} \cline{8-13} \\

 -18.9 $\geq M_{B}$  \textgreater -19.9 & 455    &19.687$\pm$0.015&4.971$\pm$0.011& 0.285  & 0.995 &&-18.9 $\geq M_{B}$  \textgreater -19.9 & 140    &19.633$\pm$0.020&4.356$\pm$0.032& 0.234 & 0.968   \\
 -19.9 $\geq M_{B}$  \textgreater -20.9 & 355    &18.714$\pm$0.020&4.953$\pm$0.016& 0.284  & 0.992 &&-19.9 $\geq M_{B}$  \textgreater -20.9 & 140     &18.910$\pm$0.024&4.356$\pm$0.032& 0.234 & 0.968  \\
 -20.9 $\geq M_{B}$  \textgreater -21.9 & 255   &17.777$\pm$0.031&4.912$\pm$0.025& 0.280 & 0.985 &&-20.9 $\geq M_{B}$  \textgreater -21.9 & 140    &18.187$\pm$0.033&4.356$\pm$0.032& 0.234 & 0.968  \\
 -21.9 $\geq M_{B}$  \textgreater -22.9 & 155   &16.978$\pm$0.060&4.786$\pm$0.048& 0.270  & 0.963 &&-21.9 $\geq M_{B}$  \textgreater -22.9 & 140    &17.464$\pm$0.045&4.356$\pm$0.032& 0.234 & 0.968  \\
 -22.9 $\geq M_{B}$  \textgreater -23.9 & 55  &16.827$\pm$0.200&4.305$\pm$0.147& 0.212  & 0.885 &&-22.9 $\geq M_{B}$  \textgreater -23.9 & 140  &16.888$\pm$0.069&4.356$\pm$0.032& 0.234 & 0.968  \\

&&&&&&&&&&&&\\

&&Case 3&&&&&&&&&\\

\cline{1-6} \\

 -18.9 $\geq M_{B}$  \textgreater -19.9 &   245 &19.702$\pm$0.018& 4.819$\pm$0.025 &0.275 &0.984\\
 -19.9 $\geq M_{B}$  \textgreater -20.9 & 194    &18.809$\pm$0.024&4.735$\pm$0.032& 0.269  & 0.976 \\
 -20.9 $\geq M_{B}$  \textgreater -21.9 & 140   &18.084$\pm$0.037&4.508$\pm$0.043& 0.253  & 0.960 \\
 -21.9 $\geq M_{B}$  \textgreater -22.9 & 85   &17.797$\pm$0.068&4.028$\pm$0.065& 0.218 & 0.925\\
 -22.9 $\geq M_{B}$  \textgreater -23.9 & 35   &18.491$\pm$0.168&3.097$\pm$0.119& 0.140  & 0.891 \\

&&&&&&&&&&&&\\

\hline

\end{tabular}
\end{minipage}

}

\end{table*}


\clearpage

\end{appendix}

\begin{appendix}


\section{Hypothesis Tests.}

\subsection{Mean value test}

Consider the case where some estimator $\hat{\Phi}$ is computed from a sample of {\it N} independent observations of a random variable {\it x(k)}. Assume that there is reason to believe that the true
parameter $\Phi$ being estimated has a specific value $\Phi_{0}$. Now, even if $\Phi$=$\Phi_{0}$, the sample value $\hat{\Phi}$ will probably not come out exactly equal to $\Phi_{0}$ because of 
the sampling variability associated with $\hat{\Phi}$. Hence the following questions arises. If it is hypothesized that $\Phi$=$\Phi_{0}$, how much difference between $\hat{\Phi}$ and $\Phi_{0}$ 
must occur before the hypothesis should be rejected as being invalid? This question can be answered in statistical terms by considering the probability of any noted difference between $\hat{\Phi}$ 
and $\Phi_{0}$ based upon the sampling distribution for $\hat{\Phi}$. If the probability of a given difference is not small, the difference would be accepted as a normal statistical variability and 
the hypothesis that $\hat{\Phi}$=$\Phi_{0}$ would be accepted.

To clarify the general technique, assume that $\hat{\Phi}$ has a probability density function of {\it p($\hat{\Phi}$)}. Now, if a hypothesis that $\Phi$=$\Phi_{0}$ is true, then {\it p($\hat{\Phi}$)}
would have a mean value of $\Phi_{0}$. The probability that $\hat{\Phi}$ would fall below the lower value $\Phi_{1-\alpha/2}$ is:

\begin{equation}
Prob[\Phi \leq \Phi_{1-\alpha/2}] \; =\, \int^{\Phi_{1-\alpha/2}}_{-\infty}\, p(\hat{\Phi})\, d\hat{\Phi} \; =\, \frac{\alpha}{2}
\end{equation}

The probability that $\hat{\Phi}$ would fall above the upper value $\Phi_{\alpha/2}$ is:

\begin{equation}
Prob[\Phi > \Phi_{\alpha/2}] \; =\, \int^{\infty}_{\Phi_{\alpha/2}}\, p(\hat{\Phi})\, d\hat{\Phi} \; =\, \frac{\alpha}{2}
\end{equation}

Hence the probability that $\hat{\Phi}$ would be outside the range between $\Phi_{1-\alpha/2}$ and $\Phi_{\alpha/2}$ is $\alpha$. Now let $\alpha$ be small so that it is very unlikely that $\hat{\Phi}$ 
would fall outside the range between $\Phi_{1-\alpha/2}$ and $\Phi_{\alpha/2}$. If a sample were collected and a value of $\hat{\Phi}$ were computed which in fact fell outside the range between $\Phi_{1-\alpha/2}$ and $\Phi_{\alpha/2}$, there would be strong reason to question the original hypothesis that $\Phi$=$\Phi_{0}$
since such a value for $\hat{\Phi}$ would be very unlikely if the hypothesis were true. Hence the hypothesis that $\Phi$=$\Phi_{0}$ would be rejected. On the other hand, if the value for $\hat{\Phi}$ fell within the range $\Phi_{1-\alpha/2}$ and $\Phi_{\alpha/2}$, there would be not strong reason to question the original 
hypothesis. Hence the hypothesis that $\Phi$=$\Phi_{0}$ would be accepted.

The small probability $\alpha$ used for the hypothesis test is called the {\it level of significance} for the test (1- $\alpha$ is called the {\it level of confidence}). The range of values of $\hat{\Phi}$ for which the hypothesis will be rejected is called the 
{\it region of rejection} or {\it critical region}. The range of values of $\hat{\Phi}$ for which the hypothesis will be accepted is called the {\it region of acceptance}. The simple hypothesis test outlined above is called a {\it two-side test} because, if the hypothesis is not true, the value of $\Phi$ could be either greater or less than $\Phi_{0}$. Hence it is necessary to test for significant differences between $\Phi$ and $\Phi_{0}$ in both directions. In other cases a one-sided test might be sufficient.

\subsection{Chi-Square test}

A special type of hypothesis test which is often used to test the equivalence of a probability density function for sampled data to some theoretical density function is called the chi-square goodness-of-fit test. The general procedure involves the use of a statistic with an approximate chi-square distribution as a measure of the discrepancy between an observed probability density function and the theoretical density function. A hypothesis of equivalence is then tested by studying the sampling distribution of this statistic.

Consider a sample of {\it N} independent observations from a random variable {\it x(k)} with a probability density function of {\it p(x)}. Let the observations be grouped into {\it K} intervals, called {\it class intervals}, which together form a {\it frequency histogram}. The number of observations falling within the
{\it i}th class intervals is called the {\it observed frequency} in the {\it i}th class, and will be denoted by ${\it f_{i}}$. The number of observations which would be expected to fall within {\it i}th class interval if the true probability density function for {\it x(k)} were ${\it p_{0}(x)}$ is called the 
{\it expected frequency} in the {\it i}th class interval, and will be denoted by ${\it F_{i}}$. Now, the discrepancy between the observed frequency and the expected frequency within each class interval is given by ${\it f_{i}}$ - ${\it F_{i}}$. To measure the total discrepancy for all class intervals, the squares of the 
discrepancies in each interval are summed to obtain the statistics

\begin{equation}
X^{2}\; =\, \sum^{K}_{i=1}\, \frac{f_{i}-F_{i}}{F_{i}} 
\end{equation}

The distribution for $X^{2}$ is approximately the chi-square ($\chi^{2}_{n}$) distribution. The number of degrees of freedom, $n$, is equal a {\it K} minus the number of different independent linear restrictions imposed to the observations. For the case where the chi-squared goodness-of-fit test is used as a test for normality, the number of degrees of freedom for $X^{2}$ is $n = K -3$.

Let it be hypothesized that the variable {\it x(k)} has a probability density function ${\it p(x)}$ = ${\it p_{0}(x)}$. After grouping the sampled observations into $K$ class intervals and computing the 
expected frequency for each interval assuming ${\it p(x)}$ = ${\it p_{0}(x)}$, compute $X^{2}$ as indicated in Eq. B.3. Since any deviation of ${\it p(x)}$ from ${\it p_{0}(x)}$ will cause $X^{2}$ increase, a one-sided test is used. The region of acceptance is:

\begin{equation}
X^{2}\; \leq\, \chi^{2}_{n;\alpha} 
\end{equation}

If the sample value of $X^{2}$ is greater than $\chi^{2}_{n;\alpha}$, the hypothesis that ${\it p(x)}$ = ${\it p_{0}(x)}$ is rejected at the $\alpha$ level of significance. If $X^{2}$ is less than or equal to $\chi^{2}_{n;\alpha}$, the hypothesis is accepted.

\subsection{Run test}

The run-test is a statistical non-parametric test which does not presuppose a specific distribution for the data which will be evaluated. A run is defined as a sequence of identical observations that are followed or preceded by a different observation or no observation at all. The number of runs which occur in a sequence of observations gives an indication as to whether or not results
are independent random observations of the same random variable.

A run-test is used when there is reason to believe that the data present an underlying trend; that is, the probability of the same event changes from one observation to the next. The mean value and variance of the number of runs in a sequence that contains $N$ observations is given by:

\begin{equation}
\mu_{r} = \frac{N}{2}+1, 
\end{equation}

\begin{equation}
\sigma_{r}^{2}=\frac{N\left(N-2\right)}{4\left(N-1\right)}
\end{equation}

The hypothesis can be tested at any desired level of significance $\alpha$ by comparing the number of observed runs (r) to the number of runs that are contained within the interval $(r_{n;1-\alpha/2},r_{n;\alpha/2})$ where $n=\frac{N}{2}$. This number is calculated from the equation that represents the distribution of the number of runs in a sequence of $N$ observations; therefore, if we suppose that there is no underlying trend in the data we analyse, and we test this hypothesis with a level of significance $\alpha$, then, $r$ in our data must satisfy the following condition $r_{n;1-\alpha/2}\leq r\leq r_{n;\alpha/2}$; if so we say that the data present no underlying trend with a significance level of $1-\alpha$ or that the significance level of an underlying trend is equal to $\alpha$.


\end{appendix}


\begin{thebibliography}{}

\bibitem[Aguerri {\it et al.} 2005]{agr05} Aguerri, J. L. A., Iglesias-P\'aramo, J., V\'ilchez, J. M.,
Mu\~noz-Tu\~n\'on, C., \& S\'anchez-Janssen, R. 2005. ApJ, 130, 475

\bibitem[Akritas \& Bershady 1996]{akr96} Akritas, M. G., \& Bershady, M. A. 1996, ApJ, 470, 706

\bibitem[Bendat \& Piersol 1966]{ben66} Bendat J. S., \& Piersol A. G. 1966. Measurement and Analysis
of Random Data. Ed. Wiley J. \& Sons. New York.

\bibitem[Bender {\it et al.} 1992]{ben92} Bender, R., Burstein, D., \& Faber S. M. 1996, ApJ, 463. L51

\bibitem[Bernardi {\it et al.} 2003]{ber03} Bernardi, M., {\it et al}. 2003, AJ, 125, 1817

\bibitem[Bernardi {\it et al.} 2003b]{ber03b} Bernardi, M., {\it et al}. 2003b, AJ, 125, 1849

\bibitem[Bernardi {\it et al.} 2007]{ber07} Bernardi, M., {\it et al}. 2007, AJ, 133, 1741

\bibitem[Bernardi {\it et al.} 2007b]{ber07b} Bernardi, M., {\it et al}. 2007b, AJ, 133, 1954

\bibitem[Bruzual \& Charlot 2003]{bru03} Bruzual G.A., Charlot S., 2003, MNRAS, 344, 1000

\bibitem[Caon {\it et al.} 1993]{cao93} Caon, N., Capaccioli, M., \& D'Onofrio, M.  1992, MNRAS, 259, 323

\bibitem[Capaccioli {\it et al.} 1992]{cap92} Capaccioli, M., Caon, N., \& D'Onofrio, M.  1993, MNRAS, 265, 1013

\bibitem[Colberg {\it et al.} 2000]{col00} Colberg, J. M., {\it et al.}.  2000, MNRAS, 319, 209

\bibitem[Djorgovsky \& Davis 1987]{djo87} Djorgovsky, S., \& Davies, M. 1987, ApJ, 313, 59

\bibitem[D'Onofrio {\it et al.} 2006]{don06} D'Onofrio, M., Valentinuzzi, T., Secco, L., Caimmi, R., \& Bindoni, D. 2006, NewAR, 50, 447

\bibitem[Dressler {\it et al.} 1987]{dre87} Dressler, A., Lynden-Bell, D., Burstein, D., Davies, R. L., Faber, S. M., Terlevich, R.,
\& Wegner, G. 1987, ApJ, 313, 42

\bibitem[Fasano {\it et al.} 2002]{fas02} Fasano, G., Bettoni, D., Marmo, C., Pignatelli, E., Poggianti, B. M., Moles, M.,
\& Kj{\ae}rgaard, P. 2002, ASPC, 268, 361

\bibitem[Fasano {\it et al.} 2004]{fas04} Fasano, G., Varela, J., Bettoni, D., {\it et al}. 2004, Private communication.

\bibitem[Fritz {\it et al.} 2005]{fri05} Fritz, A., Ziegler, B. L., Bower, R. G., Smail, I., \& Davies, R. L. 2005, MNRAS, 358, 233

\bibitem[Fukugita {\it et al.} 1996]{fuk96} Fukugita, M., Ichikawa, T., Gunn, J. E., Doi, M., Shimasaku, K., \& Schneider D. P. 1996,
AJ, 111, 4

\bibitem[Graham \& Driver 2005]{gra05} Graham, A. W., \& Driver, S. P. 2005. PASP. 22. 118

\bibitem[Graham \& Guzm\'an]{gra03} Graham, A., \& Guzm\'an, R. 2003, AJ, 125, 2936

\bibitem[Hamabe \& Kormendy 1987]{ham87} Hamabe, M., \& Kormendy, J. 1987, In: Structure and Dynamics of Elliptical Galaxies, IAU Symp.,
No 127, p.379, ed. de Zeeuw, T., Reidel, Dordrecht

\bibitem[Hoessel {\it et al.} 1987]{hoe87} Hoessel, J. G., Oegerle, W. R., \& Schneider, D. P. 1987, AJ, 94, 11€11

\bibitem[Isobe {\it et al.} 1990]{iso90} Isobe, T., Feigelson, E. D., Akritas, M. G., \& Babu, G. J.  1990, ApJ, 364, 104

\bibitem[Jorgensen {\it et al.} 1992]{jor92} Jorgensen, I., Franx, M., \& K\ae rgaard, P. 1992, A\&AS, 95, 489.

\bibitem[Jorgensen {\it et al.} 1995]{jor95} Jorgensen, I.,  Franx, M., \& K{\ae}rgaard, p. 1995, MNRAS, 273, 1097

\bibitem[Jorgensen {\it et al.} 1996]{jor96} Jorgensen, I.,  Franx, M., \& K{\ae}rgaard, p. 1996, MNRAS, 280, 167

\bibitem[Jorgensen 1999]{jor99a} Jorgensen, I. 1999, MNRAS, 306, 607

\bibitem[Jorgensen {\it et al.} 1999]{jor99} Jorgensen, I.,  Franx, M., Hjorth, J., \& van Dokkum, P. G. 1999, MNRAS, 308, 833

\bibitem[Kelson {\it et al.} 1997]{kel97}  Kelson, D.D., van Dokkum, P. G., Franx, M., Illingworth, G. D., \& Fabricant, D.  1997, ApJ, 478, L13

\bibitem[Kj{\ae}rgaard {\it et al.} 1993]{kja93} Kj{\ae}rgaard, P., Jorgensen, I., \& and Moles, M. 1993, ApJ, 418, 617

\bibitem[Kelson {\it et al.} 2000]{kel00} Kelson, D. D., Illingworth, G. D., va Dokkum P. G., \& Franx, M. 2000, ApJ, 531, 137

\bibitem[Kormendy 1977]{kor77}  Kormendy, J. 1977, ApJ, 218, 333

\bibitem[Kormendy 1985]{kor85}  Kormendy, J. 1985, ApJ, 295, 73

\bibitem[La Barbera {\it et al.} 2000]{lab00} La Barbera, F., Busarello, G., \& Capaccioli, M. 2000, A\&A, 362, 851

\bibitem[La Barbera {\it et al.} 2003]{lab03} La Barbera, F., Busarello, G., Merluzzi, P., Massarotti, M., \& Capaccioli M. 2003, ApJ, 595, 127

\bibitem[Michard 2000]{mic00} Michard, R., 2000, A\&A, 360, 85

\bibitem[Milvang-Jensen 1997]{mil97} Milvang-Jensen, B. 1997, Master's Thesis. University of Copenhagen

\bibitem[Nigoche-Netro {\it et al.} 2007]{nig07} Nigoche-Netro, A., Moles, M., Ruelas-Mayorga, A., Franco-Balderas, A., \& Kj\ae rgaard, P. 2007,
A\&A, 472, 773

\bibitem[Nigoche-Netro 2007]{nig07b} Nigoche-Netro, A. 2007, PhD Thesis. Universidad Complutense de Madrid


\bibitem[Prugniel \& Siemen 1997]{pru97} Prugniel, P., \& Siemen, F. 1997, A\&A, 321, 111

\bibitem[Reda {\it et al.} 2004]{red04} Reda, F. M., Forbes, D. A., Beasley, M., O'Sullivan, E. J., \& Goudfrooij, P. 2004, MNRAS, 354, 851

   \bibitem[Saglia {\it et al.} 1993]{sag} Saglia, R. P., Bertschinger, E., Baggley, G., Burstein, D., Colles, M., Davis, R. L.,
   McMahan, R. K., \& Wegner, G. 1993, MNRAS, 264, 961.

\bibitem[Sandage \& Peremulter 1991]{san91} Sandage, A., \& Peremulter, J-M.  1991, ApJ, 370, 455

\bibitem[Sandage \& Lubin 2001]{san01} Sandage, A., \& Lubin, L. M.  2001, ApJ, 121, 2271

\bibitem[Schlegel {\it et al.} 1998]{sch98} Schlegel, D. J., Finkbeiner, D. P., Douglas, P., \&
    Davis, M. 1998, ApJ, 500, 25.

\bibitem[Varela 2004]{var04} Varela, J. 2004, PhD Thesis. Universidad Complutense de Madrid

\bibitem[Ziegler {\it et al.} 1999]{zie99} Ziegler, B. L., Saglia, R. P., Bender, R., Belloni, P., Greggio, L., \& Seitz, S. 1999, A\&A, 346, 13.


\end{thebibliography}
\end{document}